\documentclass{svmult}
\usepackage[utf8]{inputenc}

\usepackage{amssymb}
\usepackage{multirow}
\usepackage{listings} 
\usepackage{framed}
\usepackage{rotating}
\usepackage{adjustbox}
\usepackage{pdflscape}

\usepackage{xargs}         
\usepackage[dvipsnames]{xcolor}
\usepackage[colorinlistoftodos,prependcaption,textsize=tiny]{todonotes}
\newcommandx{\unsure}[2][1=]{\todo[linecolor=red,backgroundcolor=red!25,bordercolor=red,#1]{#2}}
\newcommandx{\change}[2][1=]{\todo[linecolor=blue,backgroundcolor=blue!25,bordercolor=blue,#1]{#2}}
\newcommandx{\info}[2][1=]{\todo[linecolor=OliveGreen,backgroundcolor=OliveGreen!25,bordercolor=OliveGreen,#1]{#2}}
\newcommandx{\improvement}[2][1=]{\todo[linecolor=Plum,backgroundcolor=Plum!25,bordercolor=Plum,#1]{#2}}
\newcommandx{\thiswillnotshow}[2][1=]{\todo[disable,#1]{#2}}

\renewcommand\thesection{\arabic{section}}

\usepackage{tocloft}
\cftsetindents{chapter}{0em}{0em}
\cftsetindents{section}{0em}{2.5em}
\cftsetindents{subsection}{0em}{2.5em}
\cftsetindents{subsubsection}{0em}{2.5em}
\usepackage{xcolor,colortbl}
\definecolor{Gray1}{gray}{0.95}
\usepackage[acronym,toc]{glossaries}
\newacronym{asm}{ASM}{Assembly}

\newacronym{bpdf}{BPDF}{Boolean Parametric Dataflow}

\newacronym{cal}{CAL}{CAL Actor Language}
\newacronym{cfdf}{CFDF}{Core-Functional Dataflow}
\newacronym{csdf}{CSDF}{Cyclo-Static Dataflow}
\newacronym{cpu}{CPU}{Central Processing Unit}

\newacronym{dpn}{DPN}{Dataflow Process Network}

\newacronym{fifo}{F\textsc{ifo}}{First-In First-Out queue}
\newacronym{fsm}{FSM}{Finite State Machine}

\newacronym{gpu}{GPU}{Graphics Processing Unit}
\newacronym{gcd}{GCD}{Greatest Common Divisor}

\newacronym{ir}{IR}{Intermediate Representation}

\newacronym{jade}{\textsc{Jade}}{Just-in-time Adaptive Decoder Engine}

\newacronym{lte}{LTE}{Long-Term Evolution}

\newacronym[longplural=Models of Computation]{moc}{MoC}{Model of Computation}
\newacronym[longplural=Multiprocessor Systems-on-Chips]{mpsoc}{MPSoC}{Multiprocessor System-on-Chip}
\newacronym{mdsdf}{MDSDF}{Multi-Dimensional \gls{sdf}}
\newacronym{mpeg}{MPEG}{Motion Picture Expert Group}

\newacronym{ofdm}{OFDM}{Orthogonal Frequency-Division Multiplexing}

\newacronym{pagpan}{P-AGPAN}{Parameterized Acyclic Pairwise Grouping of Adjacent Nodes}
\newacronym{padaf}{\textsc{p}a\textsc{d}a\textsc{f}}{Parametric Dataflow Format}
\newacronym{pe}{PE}{Processing Element}
\newacronym{pimm}{PiMM}{Parameterized and Interfaced Dataflow Meta-Model}
\newacronym{pisdf}{{\Large$\pi$}SDF}{Parameterized and Interfaced \gls{sdf}}
\newacronym{preesm}{\textsc{Preesm}}{Parallel Real-time Embedded Executives Scheduling Method}
\newacronym{psdf}{PSDF}{Parameterized \gls{sdf}}
\newacronym{psm}{PSM}{Parameterized Set of Modes}

\newacronym{rtos}{RTOS}{Real-Time Operating System}

\newacronym{sadf}{SADF}{Scenario-Aware Dataflow}
\newacronym{sdf}{SDF}{Synchronous Dataflow}
\newacronym{sdr}{SDR}{Software Defined Radio}
\newacronym{spdf}{SPDF}{Schedulable Parametric Dataflow}
\newacronym{spider}{\textsc{Spider}}{Synchronous Parameterized and Interfaced Dataflow Embedded Runtime}

\usepackage[caption=false,font=footnotesize]{subfig}

\title*{Reconfigurable and approximate computing for video coding}
\author{Francesca Palumbo and Carlo Sau}
\institute{Francesca Palumbo \at University of Sassari, Italy \\ \email{fpalumbo@uniss.it}
	\and Carlo Sau \at University of Cagliari, Italy \\ \email{carlo.sau@diee.unica.it}
}

\begin{document}

\maketitle


\section{Constraints and needs of Video Coding Systems} \label{sec:intro}
Systems perception capabilities are fundamental to make modern interconnected systems smart and, in this context, video coding is certainly a key technology.
Video coding standards have been traditionally based on monolithic specifications coupled to a fixed set of profiles, capable of implementing different sub-set of functionalities of the given reference standard. Standards and profiles typically offer different features and enable different trade-offs that make them suitable for different applications and scenarios. 
Nevertheless, keeping the pace with constantly evolving scenarios and user’s needs, for example in terms of offered quality or consumed power, is not that simple, and has led to continuous releases/updates of standards and profiles. 

Ideally designers should be allowed, at design-time, to select and combine coding tools and profiles to optimally match the given requirements, while guaranteeing that customized applications are still interoperable. Nevertheless, monolithic specifications tend to hide parallelism and the dataflow nature of video coding algorithms that, on the contrary, can be successfully exploited to guarantee efficient implementations and to play with different trade-offs \cite{BhattacharyyaEJLMR11}. 

\begin{framed}
\noindent To address these issues, related to codecs design and customization, designers have been called to conceive efficient design-time methodologies capable of managing the adaptation of video-coding technologies along time.
\end{framed}

Nowadays, systems are required to be flexible and versatile, capable of supporting multiple codecs and profiles simultaneously, and to enable dynamic reconfiguration among them \cite{BystromRKd10}. Fortunately, both standards and profiles, even migrating from one generation to another, tend to present commonalities that are exploitable both at specification level and at the implementation level to facilitate (re-)configurability.  

\begin{framed}
\noindent To address these issues, related to codecs execution, designers have been called to conceive efficient run-time methodologies capable of addressing these flexibility needs.
\end{framed}

\begin{figure}[h!t]
\centering
\includegraphics[width=\linewidth]{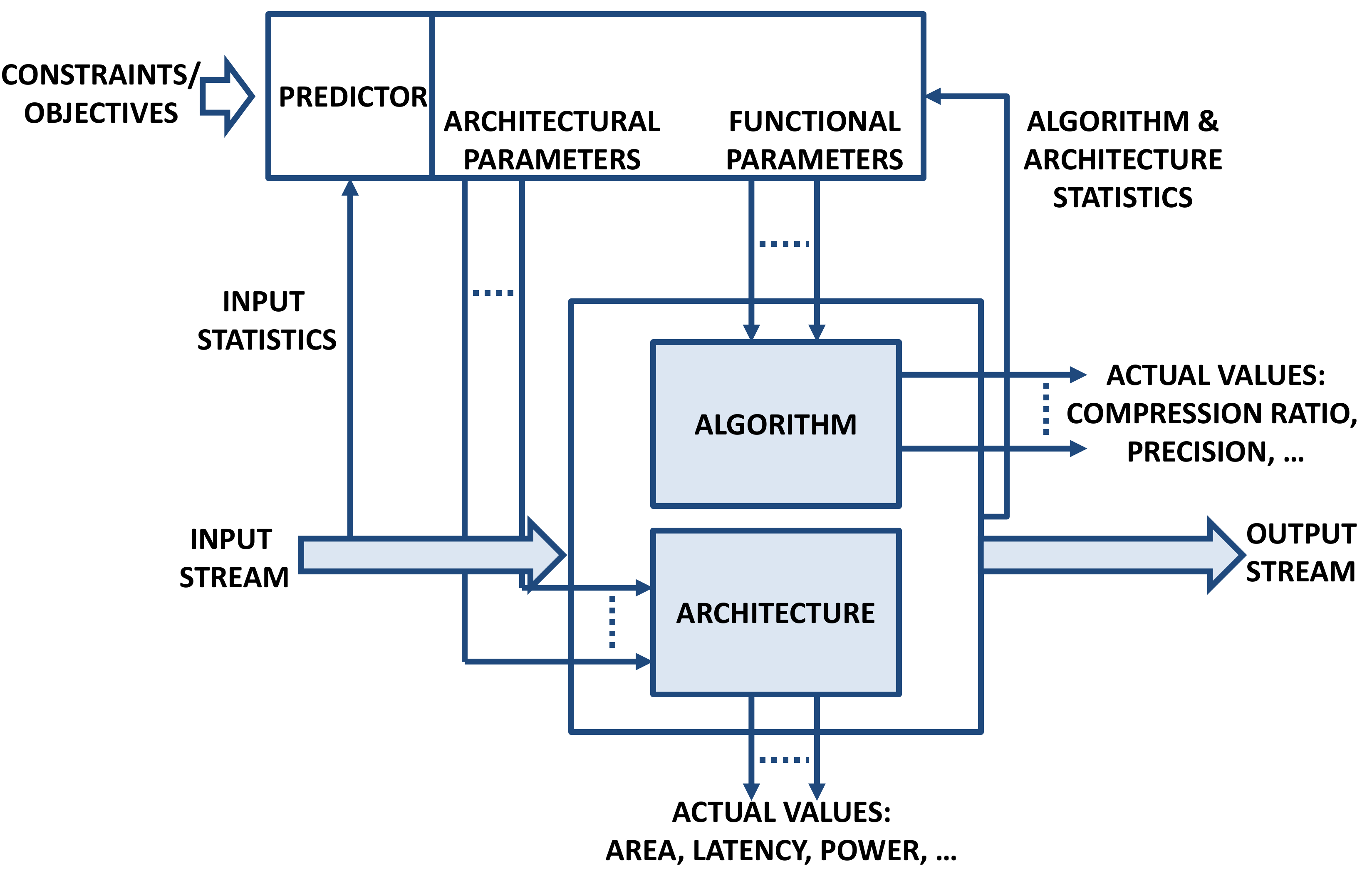}
\caption{Dynamic Parameter Adjustment \cite{BurlesonTGSJEVT01}.}
\label{fig:dpa}
\end{figure}
In \cite{BurlesonTGSJEVT01} the concept of dynamic parameter adjustment is expressed. As depicted in Figure \ref{fig:dpa}, tuning current processing to the content variation and changing user/system requirements is made possible mastering the run-time variation of different parameters, namely:
\begin{itemize}
    \item Functional Parameters - Allow for tuning output of the computation, they may include for example filter and transform lengths, and quantization levels.
    \item Architectural Parameters - Allow for tuning the guaranteed performance and energy consumption, without modifying the output bits of the computing mechanism. An example can be the computation parallelism that may affect the given throughput positively/negatively affecting the energy consumption.
\end{itemize}
\noindent Both functional and architectural parameters can be tuned to optimize codecs before deployment, and to make it also flexible to changing conditions at run-time. The aim of this chapter is describing different design-time and run-time technologies that can be adopted for these purposes.

\subsection{Chapter Organization}
The chapter is organized as follow.
\begin{itemize}
    \item Section \ref{sec:rvc}: The lack in flexibility of traditional monolithic codec specifications, not suitable to model commonalities among codecs and foster re-usability among successive codec generations/updates, was the main trigger for the development of a new standard initiative within the ISO/IEC MPEG committee, called Reconfigurable Video Coding (RVC). MPEG-RVC framework exploits the dataflow nature behind video coding to foster flexible and reconfigurable codec design, as well as to support dynamic reconfiguration.
    \item Section \ref{sec:approx}: The inherent resiliency of various functional blocks (like motion estimation in the High Efficiency Video Coding, HEVC) and the varying levels of user perception make video coding suitable to apply approximate computing techniques. Approximate computing, if properly supported at design-time, allows achieving run-time trade-offs, representing a new direction in hardware–software co-design research. The main assumption behind approximate computing, exploited within video coding, is that the degree of accuracy (in this case during codec execution) is not required to be the same all the time.
    \item Section \ref{sec:fin}: This section will try to put together the concepts addressed in the chapter and will remark on which are, on our opinion, some interesting research directions provided the open challenges.
\end{itemize}
As a final guideline in the study of this chapter, please note that for each discussed techniques emphasis on design-time and run-time support available to designers is provided. Moreover, the main objective for each discussed technique would be the (potentially) achievable run-time flexibility, which will consent them to address highly variable constraints and workloads. 

\section{Reconfigurable Video Coding} \label{sec:rvc}
The MPEG Reconfigurable Video Coding \cite{LucarzAM09}, an ISO/IEC standard, aims at providing a framework for dynamic development, implementation, and adoption of standardized video coding solutions with features of higher flexibility and re-usability. The intent of the MPEG Reconfigurable Video Coding Framework \cite{BhattacharyyaEJLMR11} is to enable a smoother transition from one standard to another, being able to maintain an effective support of the previous ones, by exploiting the commonalities among the standards. Indeed traditional monolithic schemes and specification formalism do not allow highlighting commonalities among codecs (either at the specification level or at the implementation one). To overcome this issue, MPEG-RVC provides a high level specification formalism, based on RVC-CAL that is a subset \cite{BhattacharyyaEJLMR11} of the CAL Actor Language (CAL) normalized as a part of the RVC standard\footnote{CAL is a dataflow and actor oriented language specified within the Ptolemy project at the University of California at Berkeley \cite{cal2003}.}, constituting a starting point model for the direct software and hardware synthesis. Basically, RVC does not define a codec, but provides a way to describe codecs in a modular way (combining together Functional Units, FUs, from the modular Video Tool Library, VTL), along with a set of instruments to facilitate their analysis, customization, implementation and optimization \cite{RoquierWRJMP08,LucarzPM11,NezanSWPR12,BezatiMJ13,PalumboSR13,BrunetWBMJC14,PalumboCPMR14,FanniSRP15,Casale-BrunetBM17,BezatiBMJ17,SurianoMRJST18}. Low level details of C/C++ codec implementations are abstracted away by describing a codec as an ``abstract" RVC-CAL network of actors taken from the standard library. Modularity is the key feature that enables the possibility of designing multi-standard video decoding applications and devices by allowing software \cite{AmerLMRND09,RenJSRP14} and hardware \cite{SauP14,SauRPBBM14,SauMRPBBM16} re-use across video standards.

\subsection{Overview of the standard and its features}
The MPEG RVC framework defines two standards: 
\begin{itemize}
    \item ISO/IEC23001–4 (or MPEG–B part 4) - framework and standard languages used to describe the components of the framework.
    \item ISO/IEC23002–4 (or MPEG–C part 4) - Video Tool Library
\end{itemize}
\noindent To maximize reconfiguration, a lot of effort in the MPEG-RVC community has been spent on choosing the appropriate granularity level for the components constituting the standard library. Re-use is possible if the actors specifying codecs are not too coarse but, at the same time, they should not be too fine to lead to an explosion in terms of modules in the library to make the high-level reconfiguration process unfeasible.

\begin{figure}[h!t]
\centering
\includegraphics[width=0.9\linewidth]{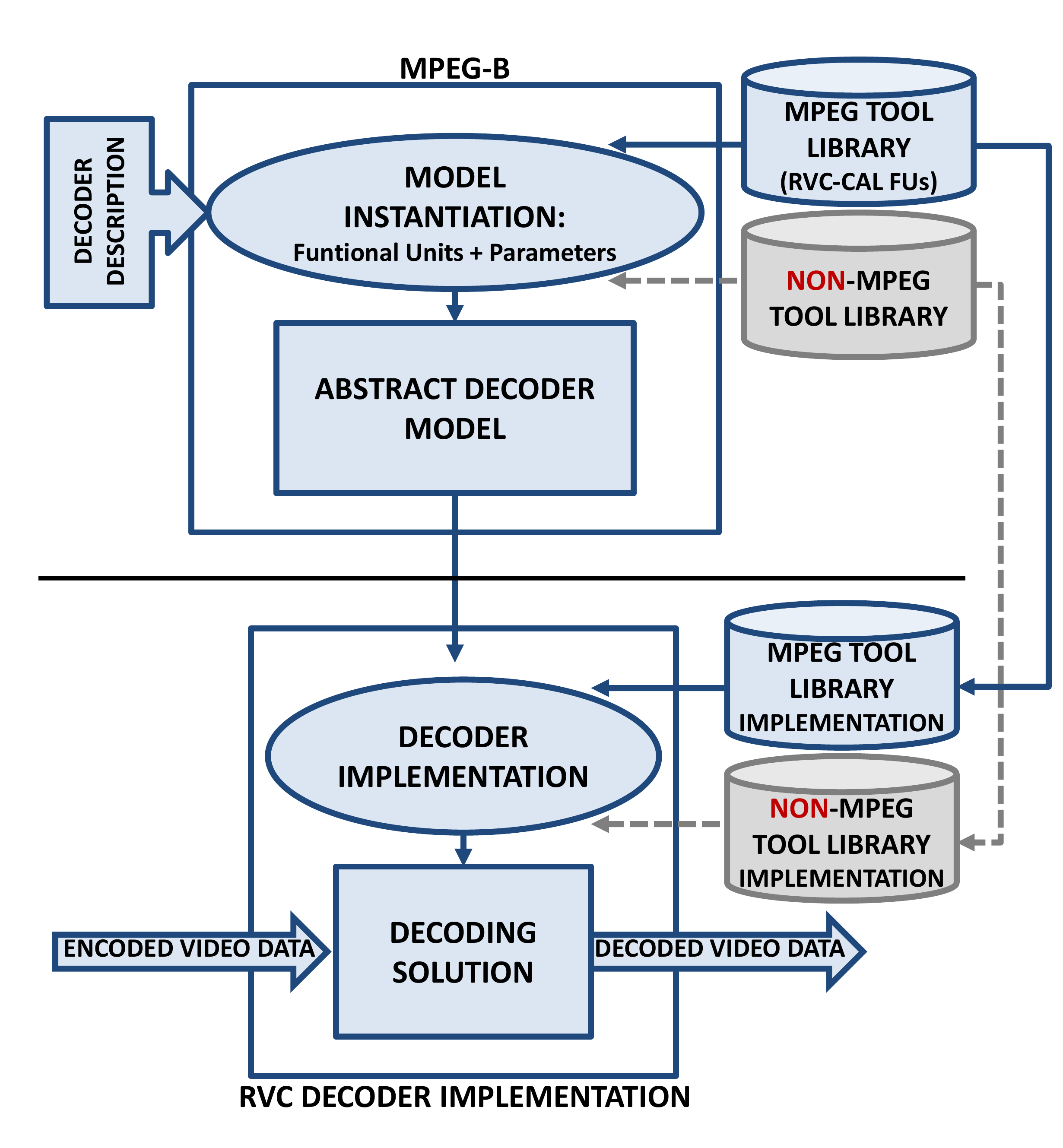}
\caption{Illustration of the RVC framework, showing the dataflow-based normative specification as well as the decoder implementation process \cite{AmerLMRND09}.}
\label{fig:rvc}
\end{figure}

The RVC-CAL features that foster its adoption in reconfigurable implementation, as reported in \cite{LucarzAM09}, are: 
\begin{itemize}
    \item natural capability of expressing parallelism - codec specifications in RVC-CAL are networks of actors that encapsulate their own state and execute concurrently. This features enables easier parallel implementations on multi-core infrastructures \cite{AmerLMRND09}.
    \item modularity - changing the implementation of an actor does not affect the rest of the network thanks to encapsulation. The dataflow form of the standard RVC specification guarantees that any reconfiguration of the user-defined proprietary library would not affect the implementation, which will remain consistent with the abstract RVC decoder model specified with the standard library.
    \item facilitated scheduling - since control and processing are decoupled. This not only allows for more facilitated scheduling, but it also allows for more flexibility. Various scheduling schemes, depending on optimization criteria, can be adopted.
    \item adaptivity - atomic firing of an action triggers the actor in a well-defined, and always known, state. One of the key features of a dataflow programming in general lays in its ability to be reconfigured dynamically.
\end{itemize}
\noindent Figure \ref{fig:rvc} illustrates both normative and proprietary components and how codecs can be built in an MPEG-RVC compliant manner. 

\subsection{Dynamic Reconfiguration Exploiting MPEG-RVC}
In terms of implementation, MPEG-RVC proved to be extremely flexible, and many different targets have been used for implementation along time. The work in \cite{AmerLMRND09} presents the automatic synthesis (C/C++ code have been synthesized leveraging on \cite{RoquierWRJMP08}) of an MPEG-4 simple profile decoder onto a dual-core platform. This work, despite not specifically reporting on a reconfigurable implementation, is particularly useful to understand the methodology for developing RVC dataflow codec and their implementation on multi-core infrastructures.

\begin{figure}[h!t]
\centering
\includegraphics[width=\linewidth]{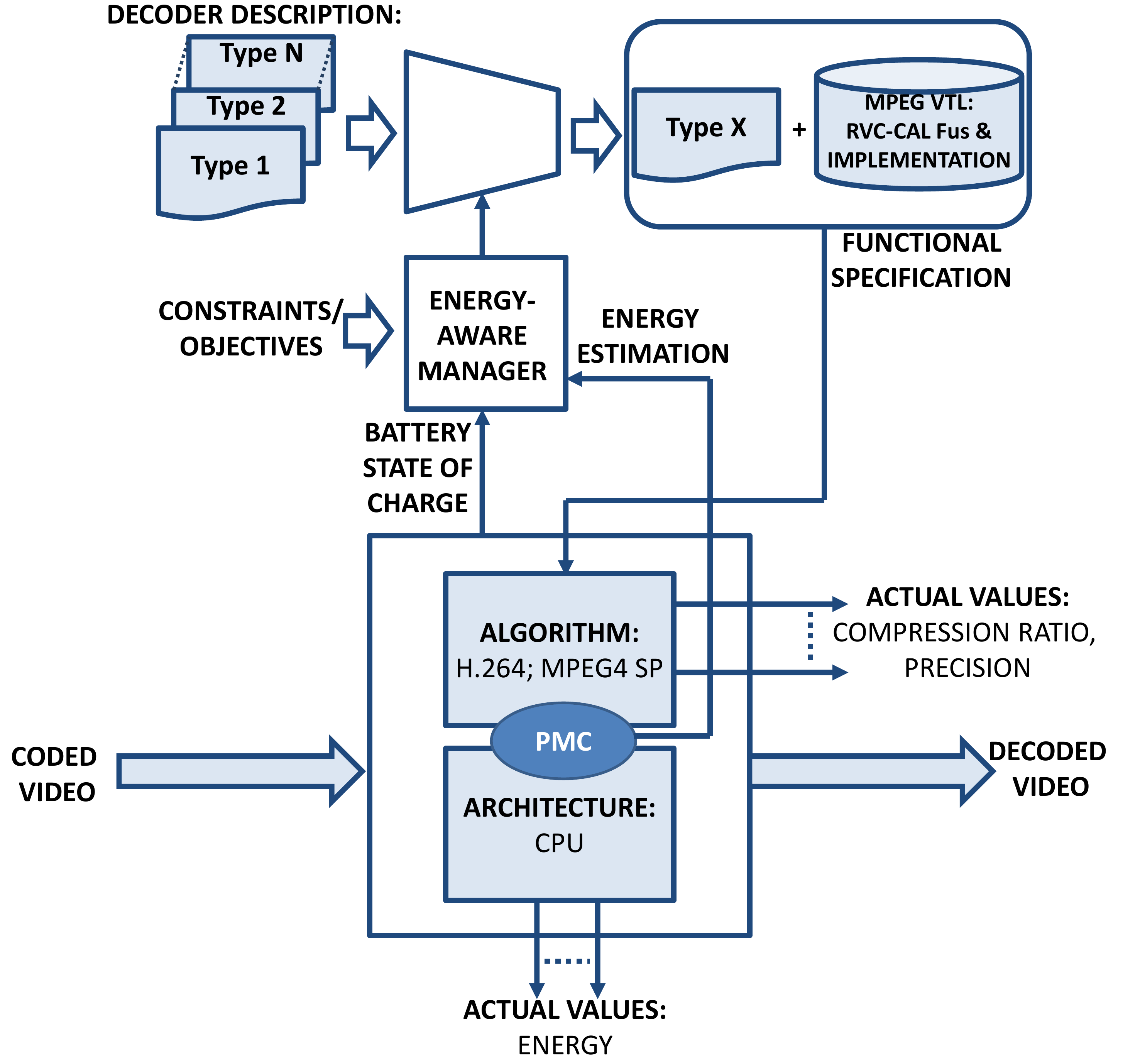}
\caption{Energy-aware decoder reconfiguration.}
\label{fig:ren}
\end{figure}
In \cite{GorinYPR11}, Gorin et al. propose the Just-In-Time Adaptive Decoder Engine (Jade) that is responsible of managing the runtime reconfiguration of a decoder, by partially/completely switching the executing network leveraging on a low level virtual machine. Reconfiguration, as discussed in Section \ref{sec:intro}, can take place for both functional and extra-functional (i.e. changes in requirements) reasons, and the work in \cite{RenJSRP14} presents a combination of Jade with on-line energy estimations to decide upon optimal decoder implementation, where the functional parameters adjusted is the decoder type itself and the information used to trigger this kind of reconfiguration is the on-line estimation of the energy consumed. In that work, Ren et al. present a Performance Monitoring Counters (PMCs) based estimation model, integrated with Jade. The model is driven by runtime retrieved counters values monitoring energy-related events. In modern processors, a set of PMCs is provided and can be accessed by high-level tools as the performance application programming interface (PAPI). In \cite{RenJSRP14} the PAPI functions are merged into the RVC-CAL framework and estimations are performed during the execution to feed Jade. Figure \ref{fig:ren} illustrates how dynamic adjustment takes place according to this technique. PMCs, constraints and current battery status are used by the \emph{energy-aware manager}; this latter is responsible of selecting which type of decoder has to be used. On-line monitoring of the battery state of charge and the power consumption of the current description of the decoder allows, once the remnant battery is lower than a pre-defined threshold, opting for a less energy consuming description.

\begin{figure}[h!t]
	\centering	\includegraphics[width=\linewidth]{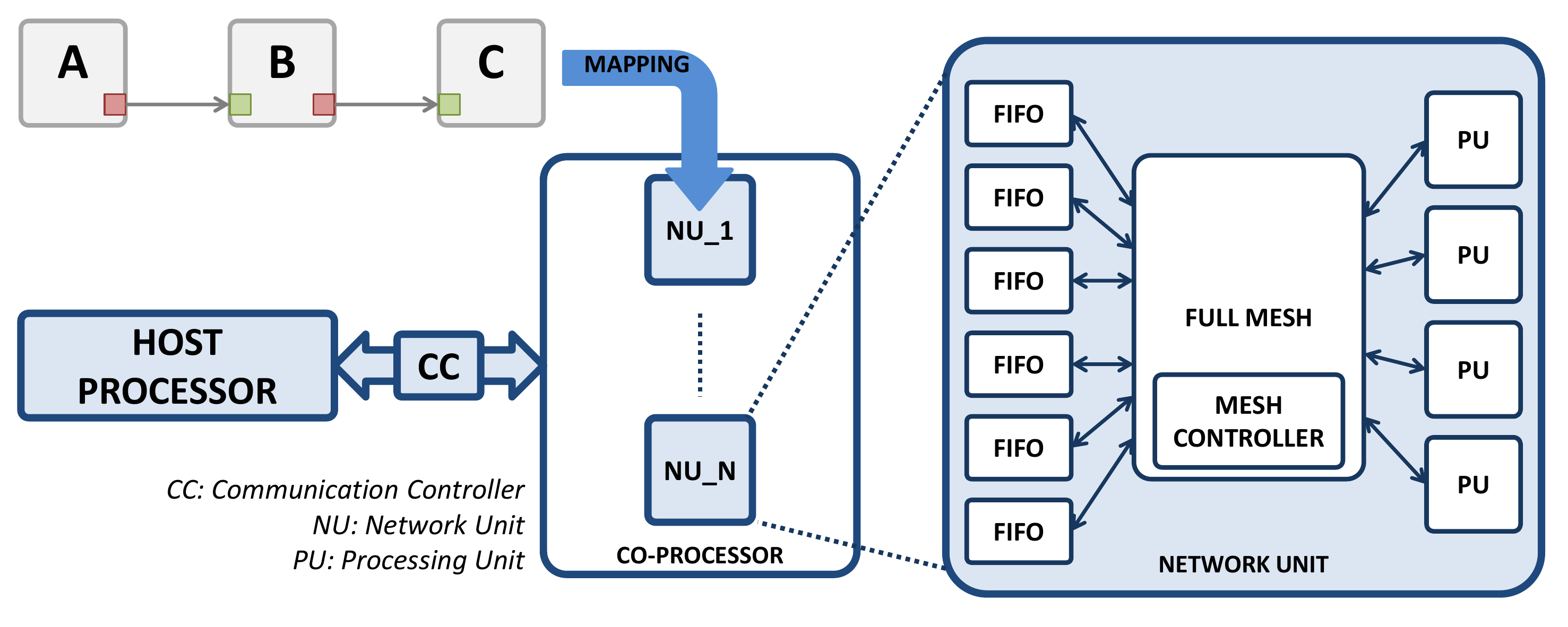}
	\caption{Dataflow Process Network based co-processor~\cite{Beaumin_2009}.}
	\label{fig:beaumCopr}
\end{figure}

\begin{figure}[h!t]
	\centering	\includegraphics[width=\linewidth]{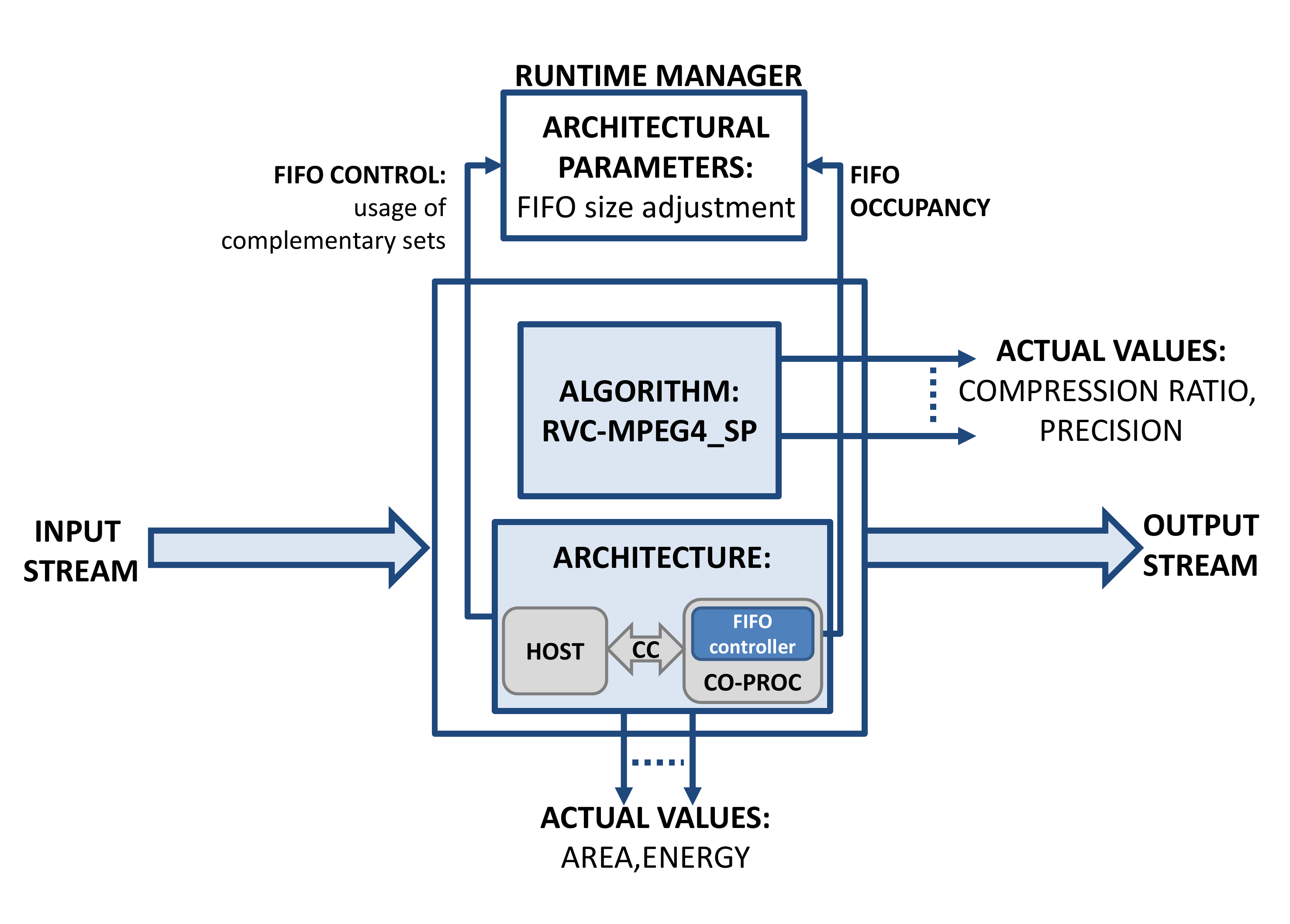}
	\caption{MPEG-RVC Reconfigurable MPEG4 SP processor to co-processor system.}
	\label{fig:beaumSys}
\end{figure}

The work of Beaumin et al. on multi-context accelerator~\cite{Beaumin_2009} can be considered the first attempt to combine dataflow specifications and reconfigurable computing, to define a co-processing unit. This latter exploits RVC-CAL principles to build the reconfigurable co-processor, which is in charge of executing different dataflow specifications. The design of the reconfigurable co-processor is based on a set of heterogeneous Processing Elements, called \emph{Network Units} (NUs). Each NU executes a different dataflow network instantiating as many \emph{processing units} (PUs) as needed to implement the CAL actors in the corresponding dataflow. Communication among PUs is FIFO-based, a full mesh infrastructure is made available in each NU and it is configurable to process any input CAL specification. An example of NU is depicted in Figure~\ref{fig:beaumCopr}. NUs are customized at design time on the basis of the input specifications. In \cite{Beaumin_2009} authors present the implementation of a co-processing unit for the MPEG4 SP implemented using the MPEG-RVC standard. The definition of the co-processing unit, with NUs and PEs, is performed at design-time. At run-time FIFO management is supported (as depicted in Figure \ref{fig:beaumSys}): if the \emph{FIFO controller} detects that the allocated memory for a FIFO is not sufficient, a complementary set of unused allocated FIFOs is activated. The required reconfiguration time is not affecting computation, since this type of reconfiguration acts at a coarse level, involving just the interconnect, and not at the level of the PUs execution. Dynamic management allows to better fit to workload variability. 

\begin{figure}[h!t]
\centering
\includegraphics[width=\textwidth]{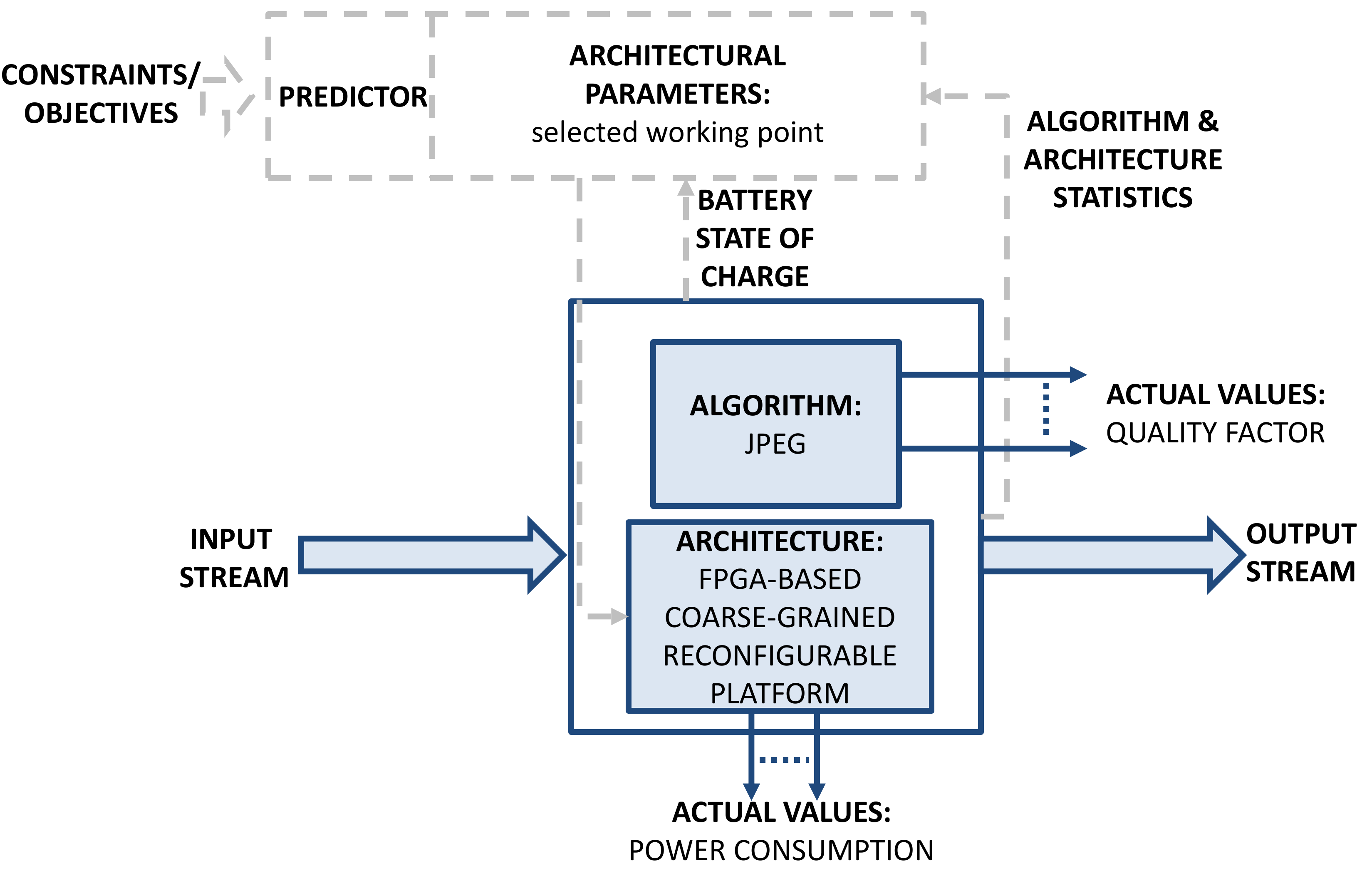}
\caption{Energy-aware decoder reconfiguration.}
\label{fig:multiQ}
\end{figure}

In \cite{SauRPBBM14} and \cite{SauMRPBBM16} MPEG-RVC is exploited for hardware reconfiguration purposes: optimized multi-functional coarse-grained virtual reconfigurable platforms are derived from disjoined specifications, exploiting the combination of three different MPEG-RVC compliant tools, namely: TURNUS \cite{turnus,BrunetWBMJC14}, Xronos \cite{BezatiMJ13} and MDC \cite{PalumboCPMR14,PalumboSFMR16}. Authors showed how coarse-grained virtual reconfiguration can be exploited to achieve functional adaptivity by implementing on a Xilinx Virtex 5 FPGA: 
\begin{itemize}
    \item an MPEG-4 multi-dataflow decoder \cite{SauRPBBM14,SauMRPBBM16} capable of switching among the MPEG-4 Simple Profile (SP) Intra-only decoder and the MPEG-4 SP Inter-decoder.
    \item  JPEG codec prototype \cite{SauMRPBBM16} capable of performing both coding and decoding over the same substrate. 
\end{itemize}
\noindent Moreover, the same coarse-grained hardware reconfiguration approach demonstrated to be effective also for requirements-oriented adaptivity purposes. In \cite{SauMRPBBM16} authors also present a multi-quality JPEG encoder, which can be depicted as in Figure \ref{fig:multiQ}. The computing part of this decoder, leveraging on coarse-grained virtual reconfiguration, enables surfing among different quality factors for the compressed images at the price of power consumption. In fact, dynamic power tends to increase going from a certain working point to better quality one, since more actors are active in the computation. Using such an approach user-triggered adaption along with power vs. performance (self-)adaptation are both feasible. The dashed parts of the figure indicate that, despite runtime reconfiguration is physically possible, the loop of dynamic (self-)adaptation was not supported at the time of that publication. 

\subsection{MPEG-RVC key remarks}\label{ss:RVCrem}
This section recaps goals and achievements of MPEG-RVC with respect to flexibility support in the video coding domain. MPEG-RVC provides a standard to define codec specification in a modular way, favoring reconfiguration in different ways as summarized hereafter:
\begin{itemize}
    \item Modular descriptions allow limiting the effort of passing from one profile to another or from one version of the standard to the successive one. Indeed, variations can be confined within atomic actors, while maintaining unchanged the rest of the network.
    \item Several different tools have been proposed at the state of the art to improved design productivity, ranging from high-level design space exploration down to hardware synthesis.
    \item Reconfiguration is also enabled at run-time and performance-/requirements-aware dynamic system management is enabled. Table \ref{tab:RVC_dynClass} provides a summary of the analyzed approaches.
\end{itemize}

\begin{table}[h!t]
\caption{Summary of Run-Time Reconfigurable RVC-based Approaches.}
\label{tab:RVC_dynClass}
\centering
\begin{tabular}{r||c|c|c|c}
\textbf{Work} & \emph{Algorithm} & \emph{Architecture} & \emph{Functional Param.} & \emph{Architectural Param.}\\ \hline 
\hline
\multirow{2}{*}{\textbf{\cite{RenJSRP14}}} & H.264 & \multirow{2}{*}{CPU} & \multirow{2}{*}{Decoder Type} & \multirow{2}{*}{---}\\  
& MPEG4-SP & & & \\ \hline 
\multirow{2}{*}{\textbf{\cite{Beaumin_2009}}} & \multirow{2}{*}{MPEG4-SP} & Host + & \multirow{2}{*}{---} & \multirow{2}{*}{FIFO depth}\\
& & Custom Co-Processor & & \\\hline
\textbf{\cite{SauRPBBM14,SauMRPBBM16}} & MPEG4-SP & \multirow{2}{*}{Custom HW} & Execution Profile & \multirow{2}{*}{Profile ID}\\ 
\textbf{\cite{SauMRPBBM16}} & JPEG & & Offered Quality & \\ \hline
\end{tabular}
\end{table}

\section{Approximate Video Coding Systems}\label{sec:approx}
Human perceptual tolerance makes video coding particularly suitable to the usage of approximate computing techniques, due to the fact that it can tolerate a varying degree of errors in the output visual quality after the coding process. 
Moreover, video coding applications approximate by construction a video coding sequence exploiting information redundancies, such as perceptual, statistical, spatial and temporal ones. For instance temporal redundancy is exploited in motion estimation that is one of the most computationally intensive steps in the HEVC standard \cite {JROhm}. Motion estimation exhibits a high degree of resilience for little arithmetical errors. Indeed, it performs a search for the already processed image block (reference block) that is mostly similar to the currently processed block (current block). Thus, if the chosen reference block is not the best one, the coding algorithm result will not be inconsistent. Actually, the general approach of fast motion estimation algorithms tends to reduce motion estimation complexity, leading to non-optimal results and causing minor degradation to the coding efficiency. Motion estimation is then a perfect candidate for the adoption of inexact operators as those available within the approximate computing field, and this is just one of the many examples. 

Approximate computing can be defined as \emph{``the idea that computer systems can let applications trade-off accuracy for efficiency"} \cite{approxOL} by accepting, under certain applicability conditions, inaccurate results rather than guaranteed accurate ones. To do that \emph{``techniques where the system intentionally exposes incorrectness to the application layer in return for conserving some resource"} are exploited \cite{approxOL}. Generally speaking the emphasis is on co-design the architectures, responsible for exposing accuracy-efficiency trade-offs, and the programming layer abstraction, which makes available the trade-offs to the programmers. Approximate solutions, for example, can be used to reduce computational complexity and energy consumption by exploring the trade-off between energy and quality of the system output. Given this consideration, approximation found good applicability in the video coding scenario. 

According to textbook definitions \cite{AgrawalCGGNOPSS16}, approximate computing provides three degrees of freedom by acting at different levels:
\begin{itemize}
    \item \emph{Data Level} - It consists in reducing the quality of the application in a controlled way by processing either less up-to-date data (temporal decimation), less input data (spatial decimation \cite{PalominoSSH16}), less accurate data (word-length optimization \cite{HilaireMS08}) or even corrupted data.
    \item \emph{Hardware Level} - The exactness of computation can be relaxed by acting on the underlying circuitry functionally, by adopting inexact operators \cite{ShafiqueHREH16}, and technologically, by playing with system parameters such as voltage supply \cite{EsmaeilzadehSCB12} \cite{VenkataramaniCCRR13}.
    \item \emph{Computation Level} - It corresponds to computation and algorithm modifications. It aims at approximating the processing to decrease the computational complexity \cite{PalominoSSH16,nogues2016algorithmic,SauPPHNMMR17}.
\end{itemize}
\noindent This classification is not so rigid, some works tend to aggregate different techniques, as in \cite{AgrawalCGGNOPSS16,PalominoSSH16} where they apply computational approximation (given by loop perforation \cite{Sidiroglou-DouskosMHR11}) in combination with data approximation \cite{ShafiqueZWBH12}, to achieve larger benefits. Moreover, other taxonomy and classifications are available in literature \cite{approxOL,Mittal2016,MoreauMWBACJS18}. In \cite{MoreauMWBACJS18}, for example, classification of approximate computing techniques is based on features\footnote{Classification of approximate computing techniques in details is beyond the scope of this chapter, but the authors in \cite{MoreauMWBACJS18} reports in Table 1 of their work a detailed an up-to-date taxonomy of approximate computing techniques referring to more than 40 state of the art works.}, as explained hereafter. 
\begin{itemize}
    \item \emph{Compute} vs. \emph{Data} - In compute-based techniques approximation takes place while running the program (e.g. in low supply voltage SRAM approximation occurs while accessing data, either for read or write). In data-based techniques approximation takes place even when the program is not running and acts on storage or representation of data values (e.g. in low refresh DRAM approximation may take place at any point in the lifetime of the data when stored). 
    \item \emph{Deterministic} vs. \emph{Non-Deterministic} - In the former case, for each and every input the same error is always provided (e.g. code perforation). In the latter one, given a set of inputs the occurrence of an error can only be probabilistically evaluated (e.g. synchronization elision).
    \item \emph{Coarse-Grained} vs. \emph{Fine-Grained} - These techniques have different scopes and range of applicability. In fact, in coarse-grained techniques arbitrary errors are induced while approximating an entire function or while reducing the overall data footprint. In this case the information loss increases as more data and more instructions are omitted. On the contrary, using fine-grained techniques errors are confined to the execution of a single instruction or on a single information.
\end{itemize}

To quantify the error and define the tolerance of approximation, for each approximate technique the \emph{Quality of Results} (QoR) has to be specified to define the level of acceptance of the computed results, according to user specifications and scenarios constraints. QoR is exploitable to guide compiler and run-time, in choosing the optimal degree of available approximation and/or for controlling the optimal approximate execution engine (determining the software and hardware approximation mechanisms to be used) \cite{CezeS16}. 

Approximate computing techniques can be clearly implemented either in software, at the code or data level, or directly in hardware, using for example inexact operators. Nevertheless, it is important to understand that in the definition of flexible solutions, implementing dynamic parameters adjustment, programmability and reconfigurability become fundamental whatever the chosen target platform is. In hardware, dynamism and flexibility are not straightforward and tuning approximation at run-time may be challenging due to the fact that execution efficiency (corresponding to system specialization) and flexibility (corresponding to system generality) are normally colliding features. To overcome this issue, as discussed in many examples in the following sections, in many cases run-time dynamism is achieved by adopting different flavors of the reconfigurable computing paradigm\footnote{Discussions and classification on reconfigurable approaches are beyond the scope of this chapter, details can be found in \cite{Compton_2002}}. 

\subsection{Data Level Approximation}
Data are a natural source of approximation. The usage of less or less precise data induces errors in the computation, which can still be tolerated according to the given scenario.

\begin{figure}[t!]
\centering
\includegraphics[width=0.9\textwidth]{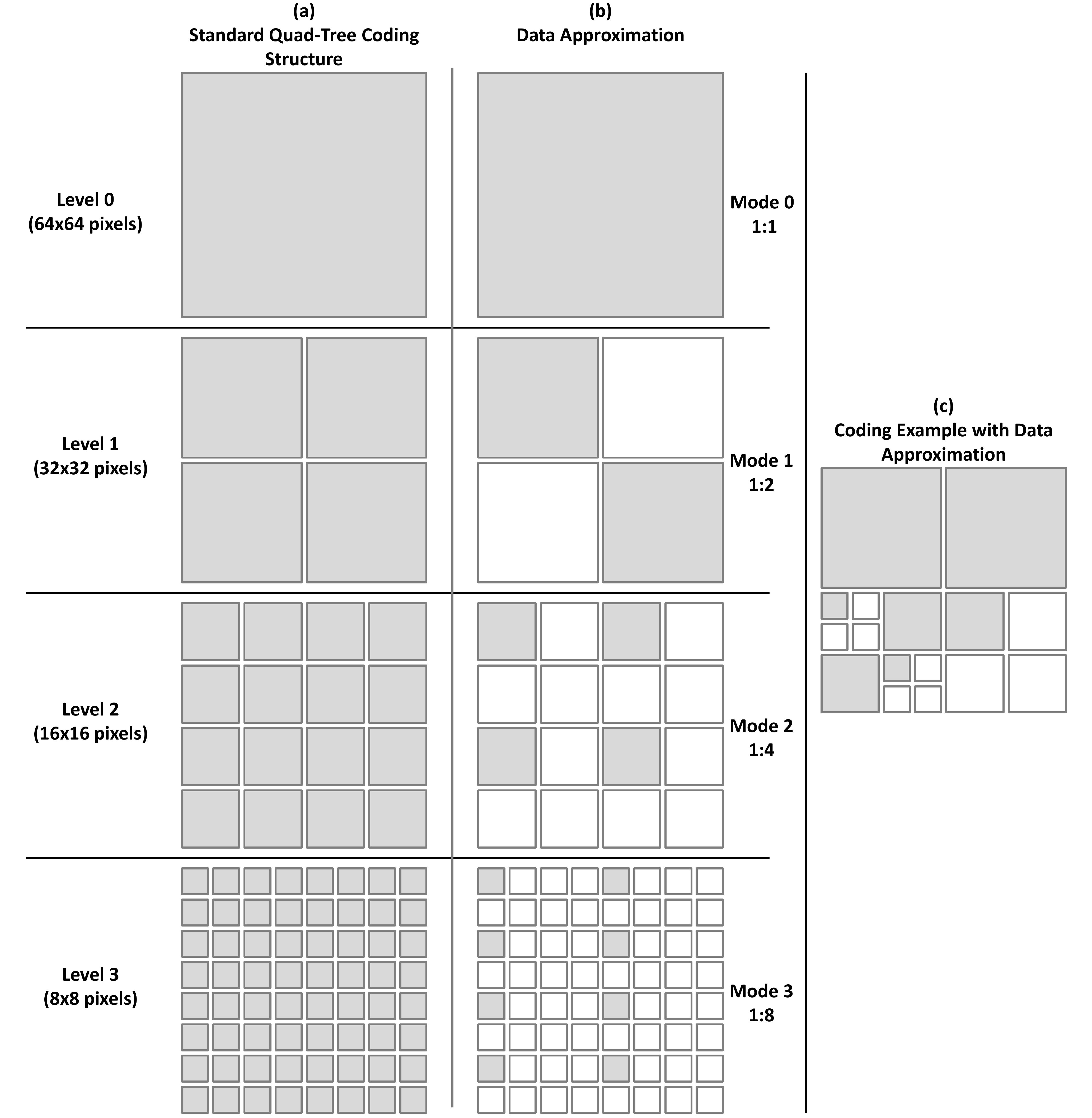}
\caption{High Efficiency Video Coding: partitioning and predictions: a) recursive quad-tree partitioning; b) data approximation [x:y means that given y pixels only x are used in prediction calculations]; c) coded frame with variable approximation degree.}
\label{fig:dataAppr}
\end{figure}

\begin{figure}[h!t]
\centering
\includegraphics[width=\linewidth]{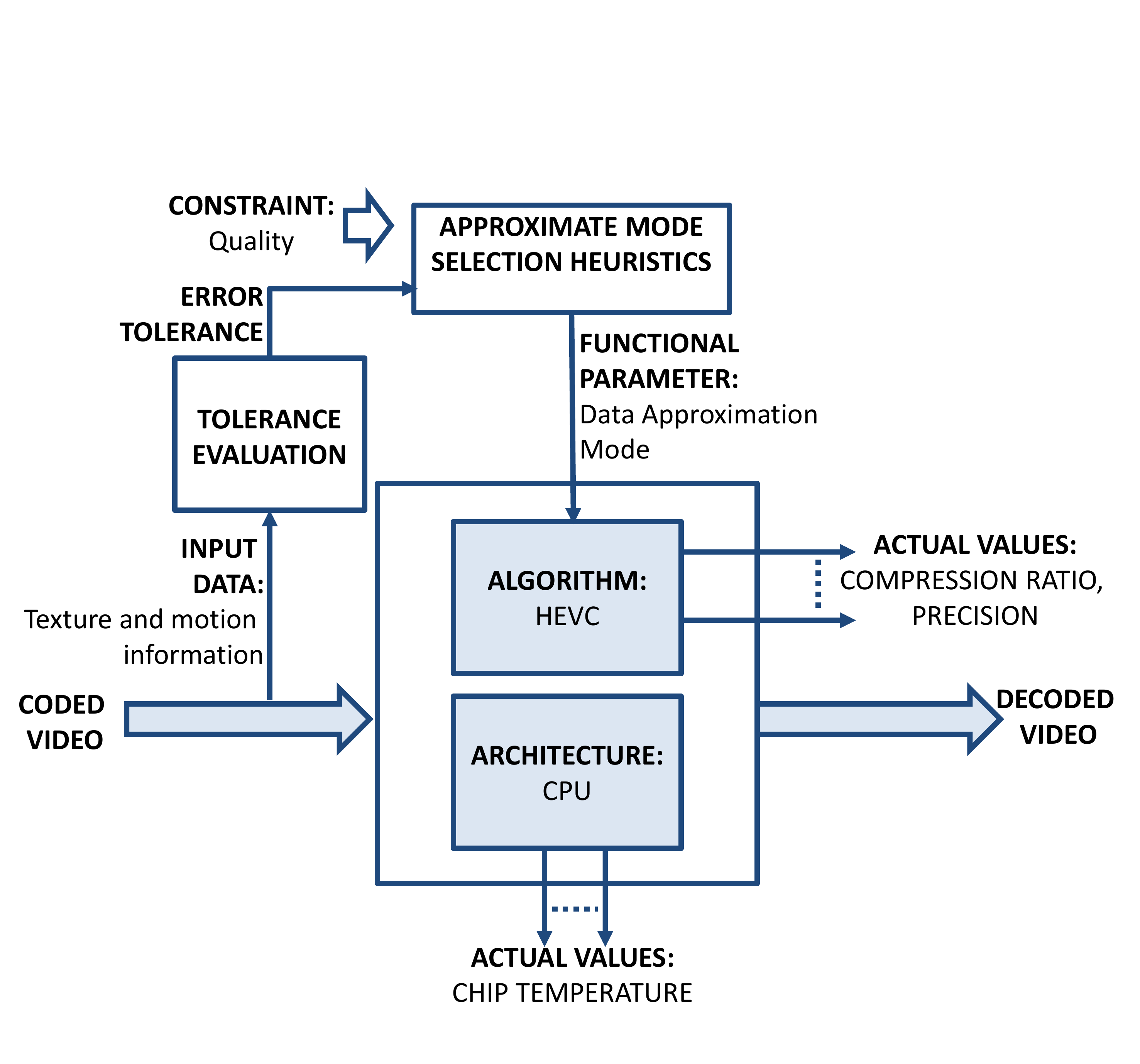}
\caption{Run-time approximate CU encoding.}
\label{fig:palomino}
\end{figure}

Palomino et al. in \cite{PalominoSSH16} presents a thermal optimization technique that employs varying degree of approximations to reduce the temperature associated with the high efficiency video coding process, while maintaining good quality results. The long term goal here is to avoid the side effects of the increased high power density of modern system on chip, where thermal hot-spots negatively affect the reliability and lifetime of devices aggravating most of the aging effects. Being video coding highly data dependent, authors adopt a content driven approximate computing technique by analyzing at run-time the tolerance to imprecise computations of each frame, to adjust the execution profile used accordingly, exploring the trade-offs between on-chip temperatures, computational complexity and visual quality. The different regions of a video sequence are classified, according to texture and motion information, to determine their potential error resilience; the higher is their tolerance supposed to be the more aggressive is the approximation (selectively eliminating operations/ computations at the algorithm and data levels) without compromising the output video quality. This work considers HEVC, which encoding process is based upon a recursive quad-tree partitioning \cite{Finkel1974} of the basic Coding Unit (CU - 64x64 pixels), which can be sub-divided iteratively till reaching 8x8 sub-blocks (see Figure \ref{fig:dataAppr}.a). Blocks are analyzed to determine the optimal compression mode. Reaching the deepest 8x8 division turned out not to be always needed: in case of homogeneous blocks (with little to none motion) 64x64 provides good coding efficiency, on the contrary highly-textured and high-motion regions require a finer splitting granularity. Spatial and temporal approximation are then used to code the sequence, leveraging on Intra- (based on matching functions like sum of absolute difference) and Inter- (based on FIR filters) predictions that are composed of primitive pixel-level operations (i.e. addition, subtraction, and multiplication), which can be approximated at various granularity achieving obviously different degradation degrees. Data-level approximation can be applied as in Figure \ref{fig:dataAppr}.b, where for each coding granularity level is presented a different data approximation degree. Figure \ref{fig:dataAppr}.c depicts an example of how variable degrees of data approximation can be used while encoding, the more aggressive is the approximation degree the more coding results may be affected. As depicted in Figure \ref{fig:palomino}, authors in \cite{PalominoSSH16} present a lightweight \emph{approximate mode selection heuristic algorithm} based on \emph{error tolerance}, which computes how far data approximation can be pushed, while preserving the requested quality. Such algorithm determines at the CU level the \emph{data approximation mode}, which is a functional parameter of quad-tree partitioning and can range from 0 (1:1, no approximation) to 3 (1:8, given 8 pixels only 1 is used in prediction calculations).

\begin{figure}[h!t]
\centering
\includegraphics[width=\linewidth]{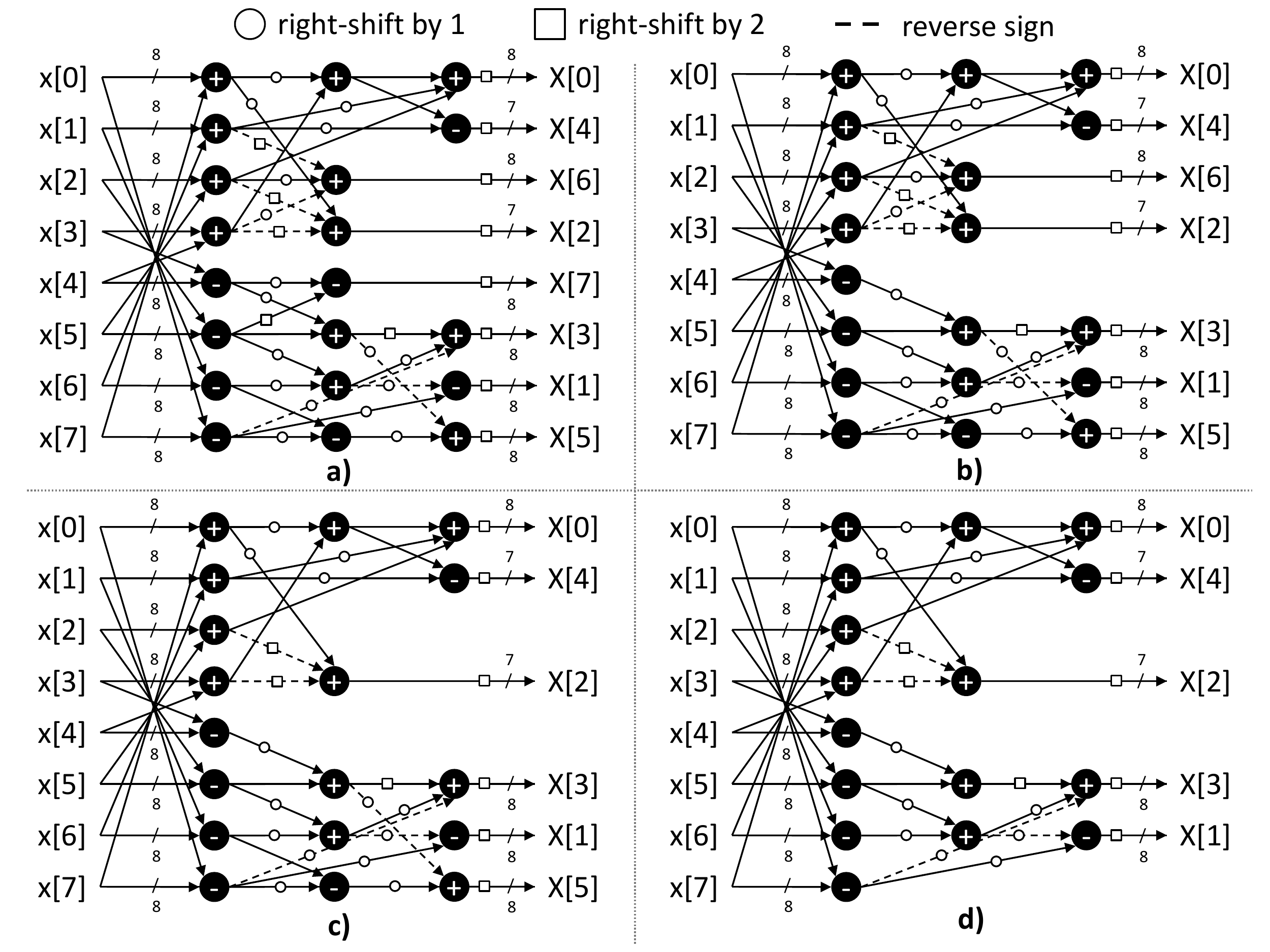}
\caption{Approximate one-dimensional DTT architecture with different numbers of calculated rows: a) 8 (all), b) 7, c) 6 and d) 5.}
\label{fig:data}
\end{figure}

Another clear example of data approximation, acting on both data precision and amount, is the work proposed by Paim et. al \cite{paim2018} on approximating the transformation step, that is typical of video coding applications. Transform operations are pervasive in video coding algorithms. They are present throughout the whole processing pipeline, including spatial and temporal prediction. Transformation means porting the input signal, in this case an image block, from the color space domain to the frequency of colors variation domain. It is implemented as a double multiplication of the image block by a given coefficients matrix, typically coming from a Discrete Cosine or Sine Transform (DCT or DST) or from alternative solutions such as the Discrete Tchebichef Transform (DTT). Paim et. al \cite{paim2018}, on the top of an already approximated (computational level approximation that will be discussed later on) version of the DTT matrix \cite{oliveira16}, apply truncation and pruning in order to simplify the computation and the system (an hardware accelerator) itself, resulting in area, power and compression efficiency improvements. Firstly, a quality-driven quantization of the DTT matrix coefficients, that is a common process in computational level approximation of transform matrices, has been performed. The chosen coefficients are quite small, making it possible to truncate up to 4 bits of the intermediate image block data and, in turn, to save lot of resources on the intermediate storage and second matrix multiplication with respect to similar literature works. Besides acting on data precision, authors also applied approximate computing on the processed amount of data: they derived three different pruned versions of the DTT accelerator by removing up to 3 rows of the considered DTT matrix (see Figure \ref{fig:data}). This led to additional improvements in terms of area, power (up to 39\% and to 43\% respectively in ASIC and FPGA with respect to the case without pruning) and throughput, in front of a quality degradation that is always less than 2 dB in terms of PSNR.

\subsection{Hardware Level Approximation}
Simplifications on the physical platform capable of performing the computation are referred to as hardware level approximation. One possibility is to simplify, or rather relax, the technological parameters of the platform: voltage supply can be lowered \cite{EsmaeilzadehSCB12} or even removed \cite{VenkataramaniCCRR13} on some or all parts of the design, always or only when approximation can be afforded. In particular, in \cite{VenkataramaniCCRR13}, the power gating applied during approximated operations is also causing data level approximation due to the resulting truncation of the operands. Another option for hardware level approximation is to adopt, by purpose, inexact operators. This means that, for instance, additions and multiplications implemented by those operators will lead to results that are not correct, but almost \cite{ShafiqueHREH16}. 

As already said, video coding is the perfect candidate application for applying approximate computing. Prediction related parts of video coding algorithms turn out to be the most suitable at this purpose, since they implement, already, approximations and they are usually the most computationally intensive steps, meaning that approximation can be very effective on them. The Sum of Absolute Differences (SAD) is the final step of the motion estimation algorithm, involved in temporal prediction. Motion estimation aims at evaluating which is the best already processed frames blocks (reference block) that is mostly similar to the currently processed frame block (current block). The decision is taken by calculating the differences between pixels of candidate reference blocks and the current one. Whereupon, the absolute values of the pixel differences are summed up with the SAD step.

\begin{figure}[t!]
\centering
\includegraphics[width=3in]{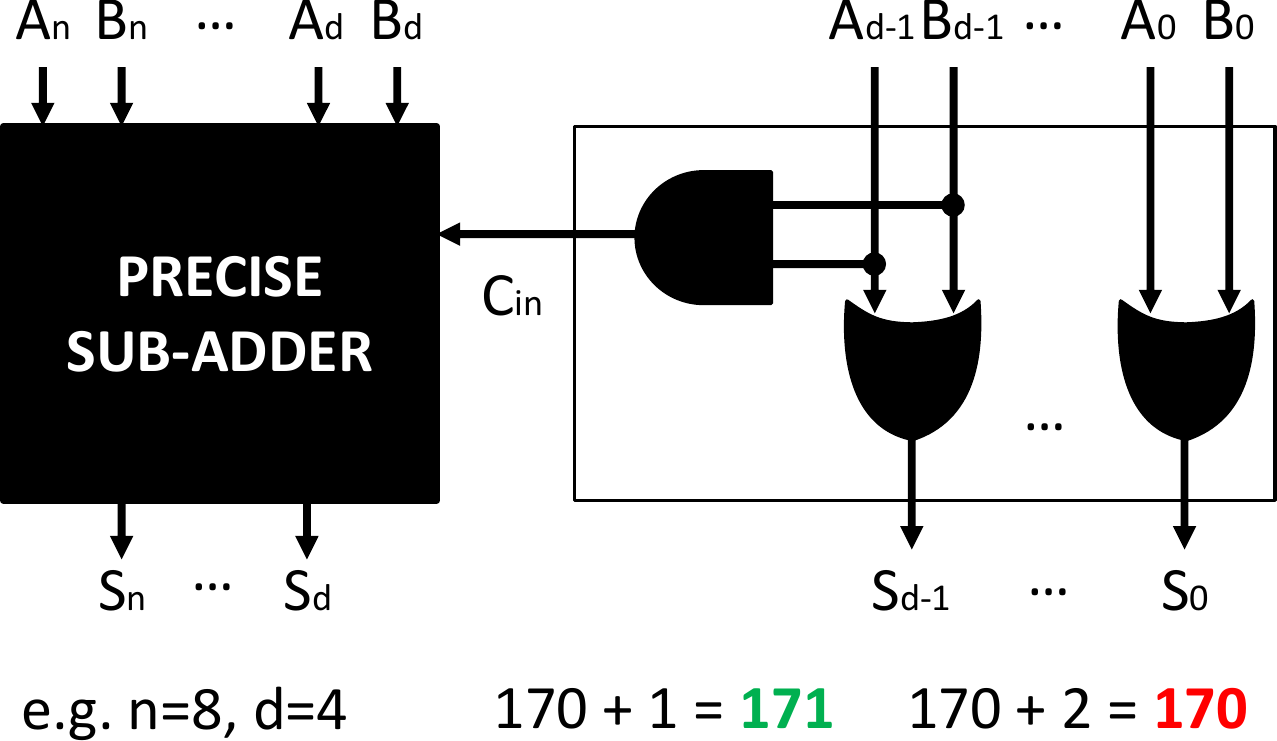}
\caption{Lower-Part-OR Adder (LOA) architecture.}
\label{fig:loa}
\end{figure}

Inexact operators, in particular adders, can be applied on SAD to simplify the process, reducing the complexity and, in turn, latency (frequency) and energy consumption \cite{porto17}. One alternative to regular, exact Ripple-Carry Adders (RCAs), commonly adopted in SAD, are Lower-Part-OR Adders (LOAs). LOAs split addition onto two pieces: the most significant bits are processed with a regular, precise sum; the least significant bits instead i) are not subjected to carry propagation; ii) are processed by means of bitwise OR operations; iii) generate carry-in for the precise portion of the design with an AND of the most significant bits of the imprecise logic (see Figure~\ref{fig:loa}). Porto et. al~\cite{porto17} studied the adoption of LOAs with different widths of the inexact portion of the operator within a hardware accelerator for motion estimation of image blocks with different sizes, up to 16x16 pixels. The hardware accelerator is strongly parallel, being able to compare the current block with 169 candidate reference blocks at a time. LOAs are applied only to the adder operators calculating the difference between pixels of the reference and current block. Overall, the core of the accelerator involves 364 adders, among which 208 have been replaced with imprecise LOAs. Three different amounts of approximated least significant bits, 3, 4 and 5, have been considered for the LOAs, resulting in three different levels of imprecision. Results show that these levels can be quantified in terms of Bjøntegaard Delta rate (BD-Rate) that is between 0.6\% and 2.5\% for the luma component and between 0.45\% and 1.85\% for the chroma one, respectively going from 3 to 5 bits on the approximated part of the adder (all these values make the quality degradation negligible). This is the price that, with respect to a regular full RCA design, has to be paid for achieving a power saving going from 9.7\% to 22.1\% on the single SAD computing core and from 7\% to 11\% on the whole accelerator.

\begin{figure}[h!t]
\centering
\includegraphics[width=\textwidth]{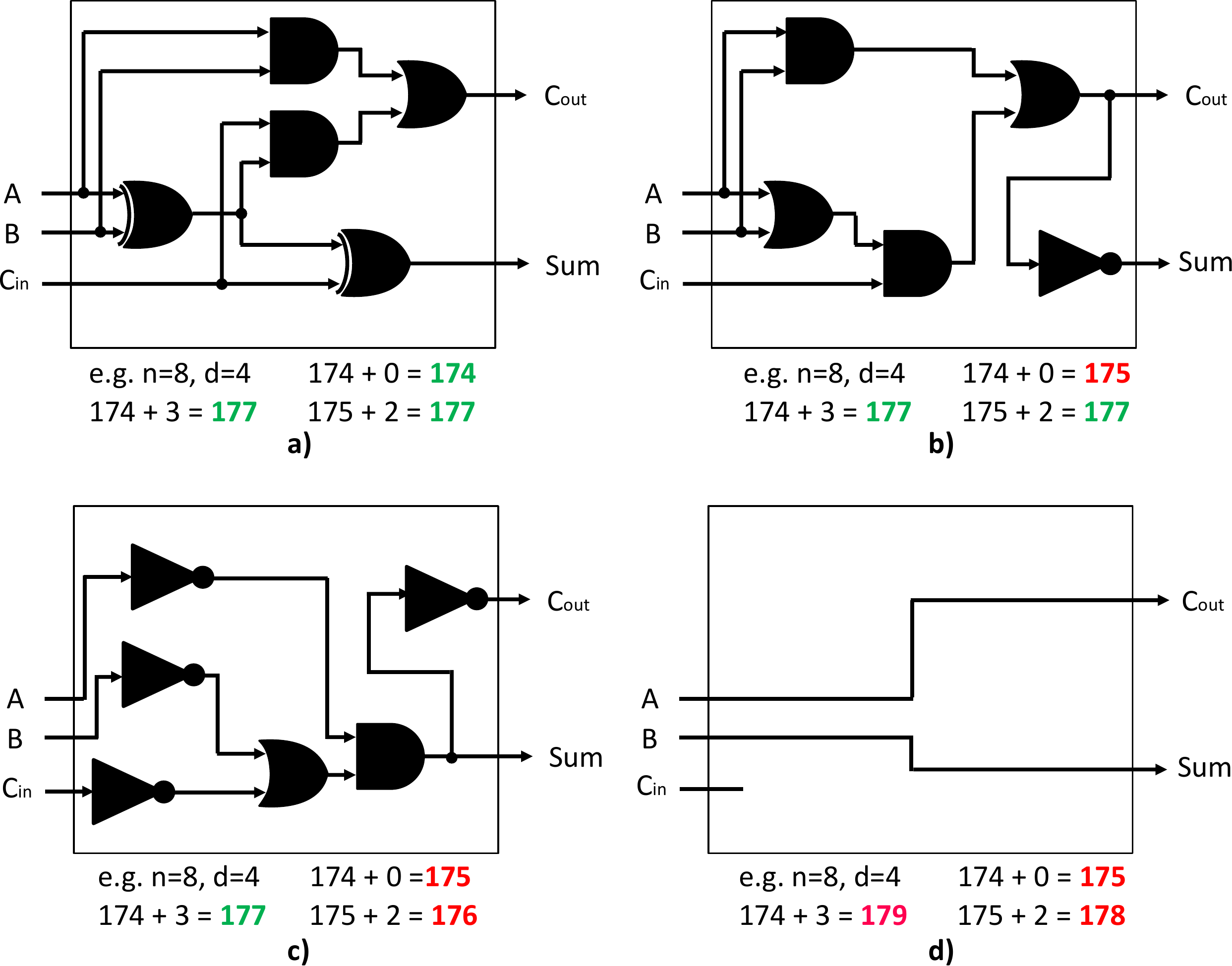}
\caption{Four different implementations of single bit adders: a) exact or b), c), d) approximate.}
\label{fig:elhar}
\end{figure}

\begin{figure}[h!t]
\centering
\includegraphics[width=\linewidth]{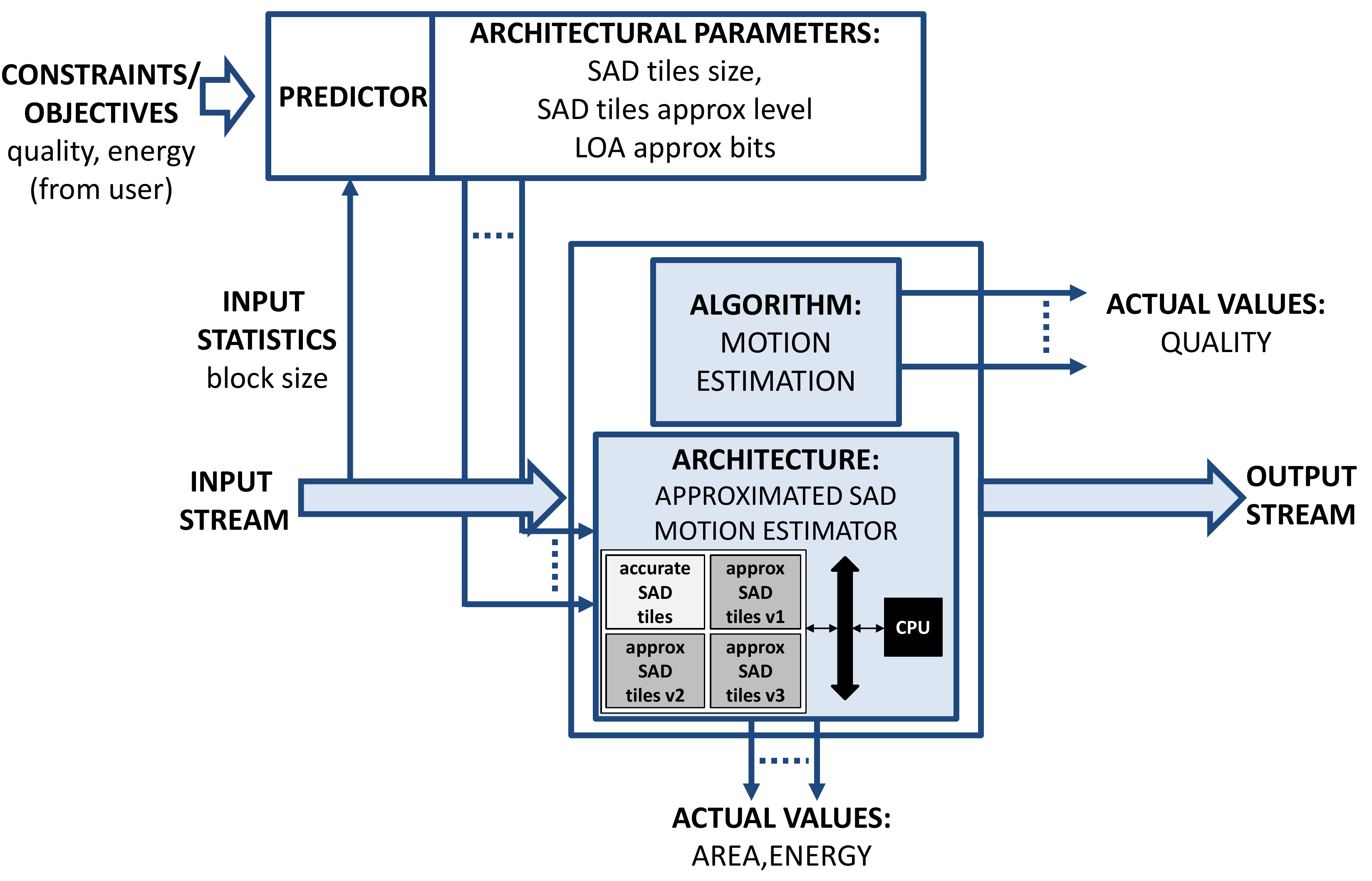}
\caption{Energy/quality/size-aware approximated motion estimation accelerator.}
\label{fig:hw_approx}
\end{figure}

El Harouni et al. \cite{elHarouni2017} also investigated the possibility of employing approximate adders on the SAD step of the video coding motion estimation, going even more deeply throughout a dynamic approximate video codec solution. Again, they adopted LOAs, but they realized four different implementations of the imprecise adders for the least significant bits, starting from the corresponding single bit instance, going from a full precise solution to a trivial only-shortcut-based one  (see Figure~\ref{fig:elhar}). LOAs have been considered to build different SAD computing tiles for several block sizes and with various bit length of the imprecise sub-adder. The SAD computing tiles have also been profiled in order to be known their bitrate, power/energy consumption and quality. An interesting measure here is that, on the overall motion estimation performance, for a given approximate adder architecture not always taking less bits on the imprecise sub-adder leads to the best results. The bitrate, power/energy and quality information is made available at run-time to a selection unit capable of decide, at run-time, which SAD tiles have to be adopted according to the current user requirements in terms of tolerable error and required energy reduction. In such a way, the system activates only the SAD tiles necessary to meet user (quality and energy) and application (block size) constraints, while power gating the others (technological hardware level approximation combined to the functional, operator related one). The resulting architecture is basically capable of tuning at run-time two kinds of architectural parameters: the SAD tiles size and their approximation level. In such a way, as can be seen in Figure \ref{fig:hw_approx}, the resulting quality (functional) and energy (architectural) values can be adjusted, according to the user requirements, as well as the specific input stream (block size impacts also on energy since only the properly SAD tiles are used in the current computation, while the others are switched off). The maximum bitrate enhancement for the final architecture is achieved for the maximum bit length of the imprecise sub-adder within the LOAs (6 bits), while in the worst case the PSNR of the decoded video is decreased by 0.57 dB. In terms of energy, the same behavior is not seen since it is not always true that the lowest consumption is reached for the most approximated LOAs.

Hardware approximation can be achieved also working on memories; this latter are one of the main sources of power dissipation in multimedia applications due to the number of memory accesses. It is possible to trade precision with consumption, lowering VDD. Supply voltage scaling is one of the most effective techniques for power reduction, as dynamic power dissipation is quadratically dependent on VDD. Nevertheless, low-voltage operations cause the failure probabilities of SRAM bit-cells to significantly increase.
In the video coding scenario this is doable, again due to the fact that human perception, which makes our eyes not so much sensitive to small variations. The work in \cite{AtaeiS18} presents 64 kB approximate SRAM architecture for low-power video applications, where 6 transistor (6T) memory cells are used. To avoid significant error this works loads the most significant bits in protected cells, and the less significant ones in unprotected cells. Three different supply voltages are enabled. Protecting the most significant bits allows for preserving a good image quality, PSNR is still around 30 db when the VDD is decreased to 0.6 V, in comparison to standard non protected ones, where PSNR is lowered down to 5 db.

\subsection{Computation Level Approximation}
Approximate computing can also be adopted at the algorithm level, in order to simplify the processing and benefit from the consequent throughput, latency, consumption improvements. Computation level approximation is basically the main pillar at the basis  of video coding algorithms by nature. However, further approximations on the computational level are still possible at different stages of the algorithm.

\begin{figure}[t!]
\centering
\includegraphics[width=\linewidth]{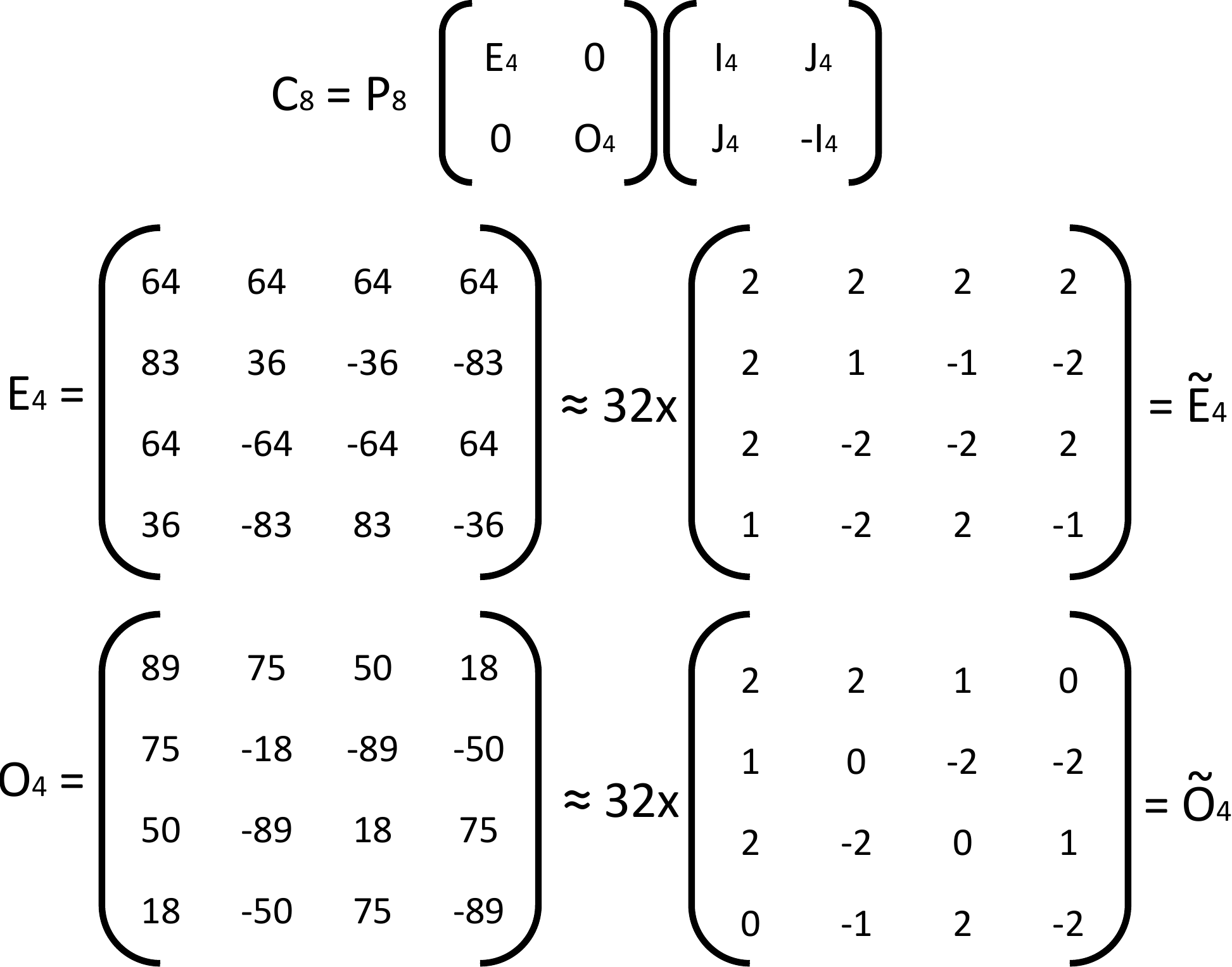}
\caption{Example of even-odd decomposed 8x8 DCT matrix approximation by quantization: $C_8$ is the original 8x8 DCT matrix, $P_8$ a 8x8 permutation matrix, $I_4$ and $J_4$ the 4x4 identity and its reflection respectively, $E_4$ and $O_4$ the even-odd 4x4 decomposition matrices that are approximated with $\widetilde{E}_4$ and $\widetilde{O}_4$ respectively.}
\label{fig:29}
\end{figure}

As already seen previously, video coding transformation step offers a certain amount of place for applying approximate computing. Actually, the coefficient matrices adopted in transformations can be seen themselves as approximations of more precise solutions \cite{gonzalez2011}. One of the most beaten research street is related to the further approximation of the video coding transform matrices. Indeed, it is very common to replace DCT matrix with a quantized DCT matrix where coefficients, that are decimal numbers lying in the range [-1,1] are rounded to the closest integer value, resulting in a Rounded Cosine Transform (RCT). In other cases the rounding is performed smartly by differentiating highest and lowest values of each matrix row from the remaining coefficients in order to preserve correlation between pixels \cite{Jridi2018} (see Figure~\ref{fig:29}). Additionally to the coefficients rounding approximation, also similarities between DCT matrices or decomposed (odd-even decomposition) DCT matrices blocks are exploited to reuse parts of the design for smaller or bigger block size transformations \cite{Jridi2017}. Decomposed DCT matrices blocks approximation to the same block has been also coupled to approximate computing at the data level: truncation of the matrix block coefficients in an odd-even decomposed matrix where odd block is used for both odd and even part has been proposed in literature \cite{renda2017}. In other cases, always relying on DCT matrix decomposition, block matrix coefficient simplification has not been driven by mathematical (simply rounding), input data (correlation between pixels) or data precision (truncation) motivations, but it has been driven by hardware implementation. Sadhvi Potluri et al. \cite{potluri2014} proposed a DCT simplified block matrix where coefficients have been approximated in order to minimize the number of multiplications and additions required for the transformation computation, resulting in a 8x8 DCT calculation with only 14 additions.


\begin{figure}[t!]
\centering
\includegraphics[width=\linewidth]{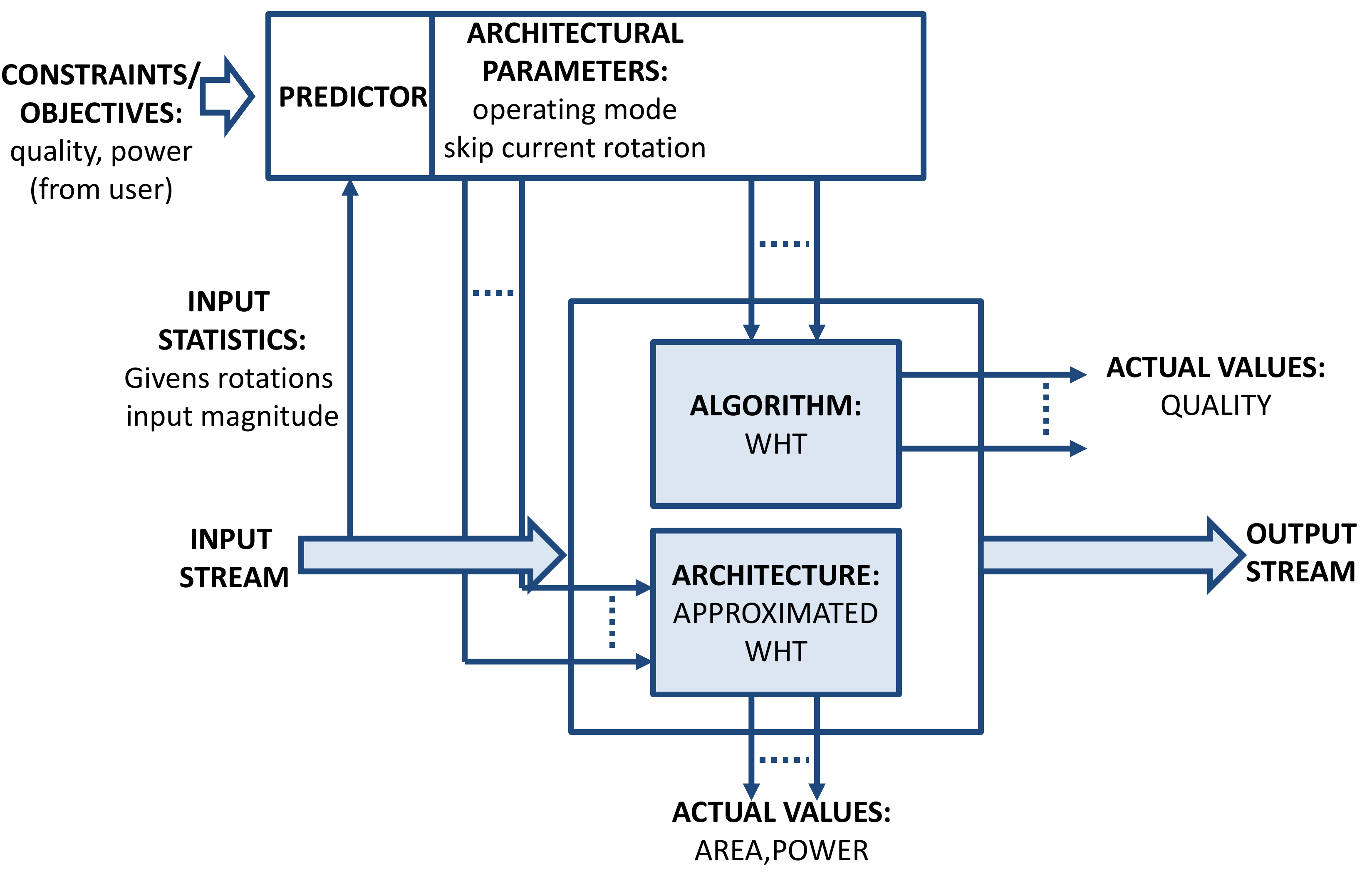}
\caption{Energy/quality approximated WHT (DCT calculation) accelerator.}
\label{fig:masera}
\end{figure}

Hardware implementation driven approximation has been also the approach followed by Masera et al. \cite{masera2017}, where authors adopted the Walsh-Hadamard Transform (WHT) to calculate the DCT. WHT is basically a decomposition of the DCT mainly requiring, besides the computing of the DCT common butterfly operations, also the calculation of the Givens rotations. If all the Givens rotations are calculated, WHT is capable of providing a complete, exact DCT. Authors proposed to skip some or all Givens rotations calculations in order to simplify the complexity of the algorithm and save power, but introducing errors in the results since WHT with less Givens rotations than expected is only an approximation of DCT. A study on how many Givens rotations have to be skipped depending on the magnitude of the DCT inputs and on the generated quality degradation has been conducted. Two different implementations have been developed, folding or not the DCT calculation circuitry for horizontal and vertical multiplication. The implementations have four different operation modes for the DCT implementation going from 0\% (no power saving, no quality degradation), to 100\% (maximum power saving, maximum quality degradation), passing by two intermediate points skipping respectively 37\% and 55\% of the Givens rotations. It is also possible to decide if Givens rotation has to be computed for a specific input by setting a certain threshold on its magnitude, thus obtaining an adaptive behavior. Such architecture can then act dynamically and change the operation mode (architectural parameter), that is power versus quality trade-off, at run-time depending on the user needs or, adaptively, on the current input data magnitude (see Figure \ref{fig:masera}). 

Generally speaking, to be effective algorithmic approximations must target computation intensive blocks. In the High Efficiency Video Coding (HEVC) standard, it has been demonstrated that computation is dominated by motion compensation \cite{BossenBSF12,nogues2016algorithmic} that is implemented with fractional pixel interpolation filters and represents between 62\% and 80\% of a decoder complexity. Basically, motion compensation leverages on temporal redundancy of the video sequences to transmit only motion vectors of the objects (image blocks) in the scene rather than complete, very similar frames. When fractional pixel motion has to be compensated, interpolation is necessary to predict a certain image block from the corresponding reference one. For instance, in HEVC luma and chroma interpolations are implemented by means of two separable Finite Input Response (FIR) filters respectively acting horizontally and vertically. 

\begin{figure}[t!]
\centering
\includegraphics[width=\linewidth]{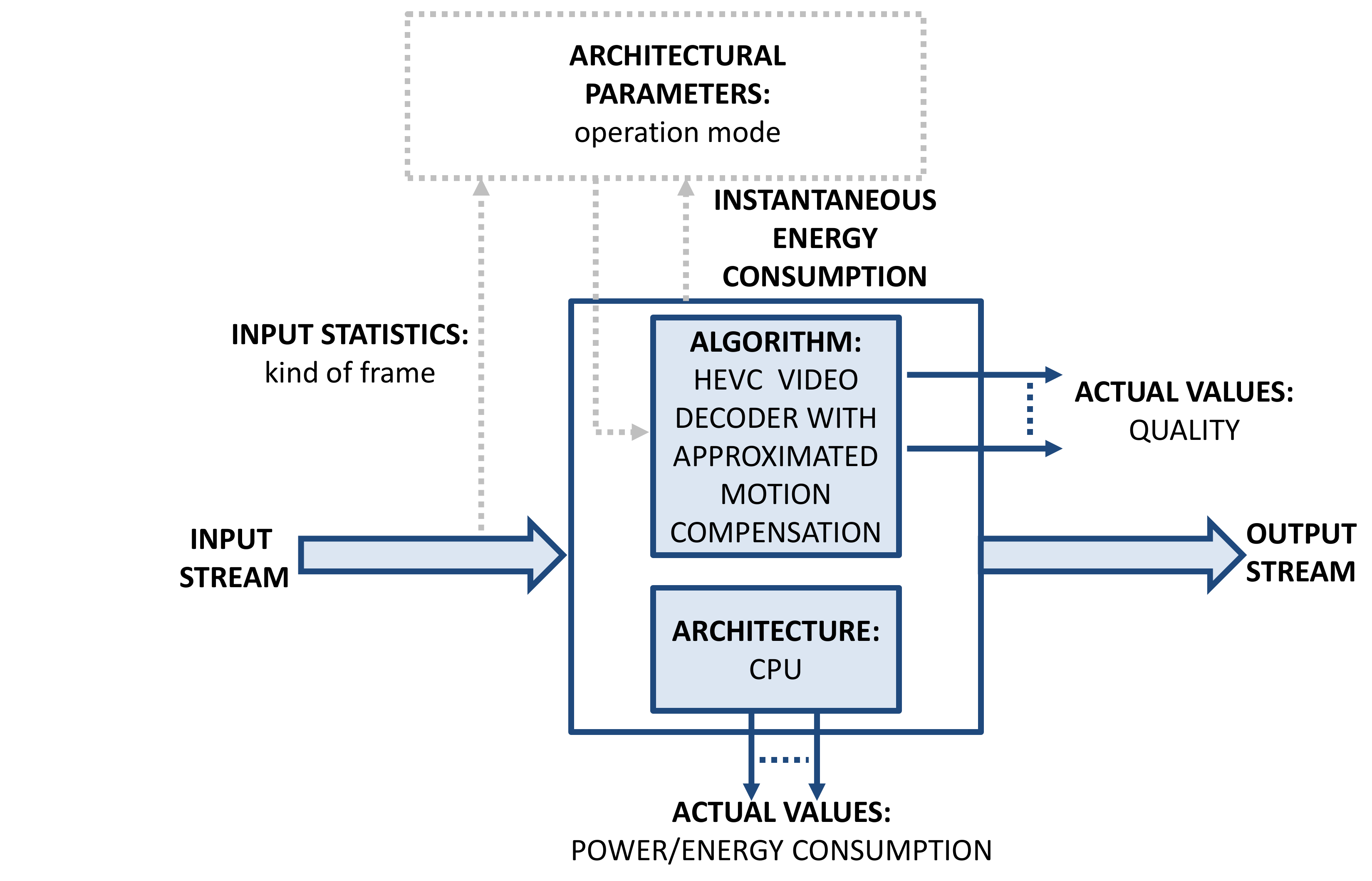}
\caption{Energy/quality approximated motion compensation for HEVC decoding.}
\label{fig:nogues}
\end{figure}

Computation level approximate computing can be applied in the motion compensation step of HEVC to reduce filtering complexity: instead of using legacy $N$ taps filters ($N$ is 7 or 8 for luma and 4 for chroma in HEVC), approximated solutions with reduced number of taps can be very effective. Up to 20\% of energy can be saved for software HEVC decoder with approximate 7, 5 and 3 taps luma, and 3 and 2 taps chroma motion compensation interpolation filter \cite{nogues2016algorithmic}. Authors combine this kind of approximation with the skipping of in-loop filtering (deblocking and SAO), constituting the final step of motion compensation in the video decoder. Four different operating modes have been proposed, each one with a different interpolation and in-loop filtering approximation and, in turn, different levels of energy savings up to 40\% of the total consumption. However, energy saving does not come for free. A price has to be paid in terms of quality of the decoded images. In particular, in the worst case the quality degradation in PSNR is about 2 dB. Note that approximation has to be applied selectively in order to avoid drawbacks on the results of the computation while, at the same time, taking advantages from its benefits in terms of energy saving. For instance, in the HEVC fractional interpolation case, approximate filters usage should be avoided on I-frames to limit quality drift between reference frames while achieving substantial energy savings \cite{nogues2015modified}. Authors envisioned also closing the loop and controlling the operating mode (quality versus energy trade-off) of the decoder according to the instantaneous energy consumption (see Figure \ref{fig:nogues}). In an extremely user oriented application field such as the video coding one, an important metric is also related to the user itself. Thus, besides the numerical, PSNR based quality degradation evaluation, the perception of the final user represents a key aspect to understand if a certain approximated implementation can be actually applied in the practice. Subjective tests have been performed for approximate HEVC fractional interpolation: the degradation of the perceptual quality seems to be imperceptible for most of the considered sample testers \cite{Sidaty17_Eusipco}. 

\begin{figure}[t!h]
\centering
\includegraphics[width=\linewidth]{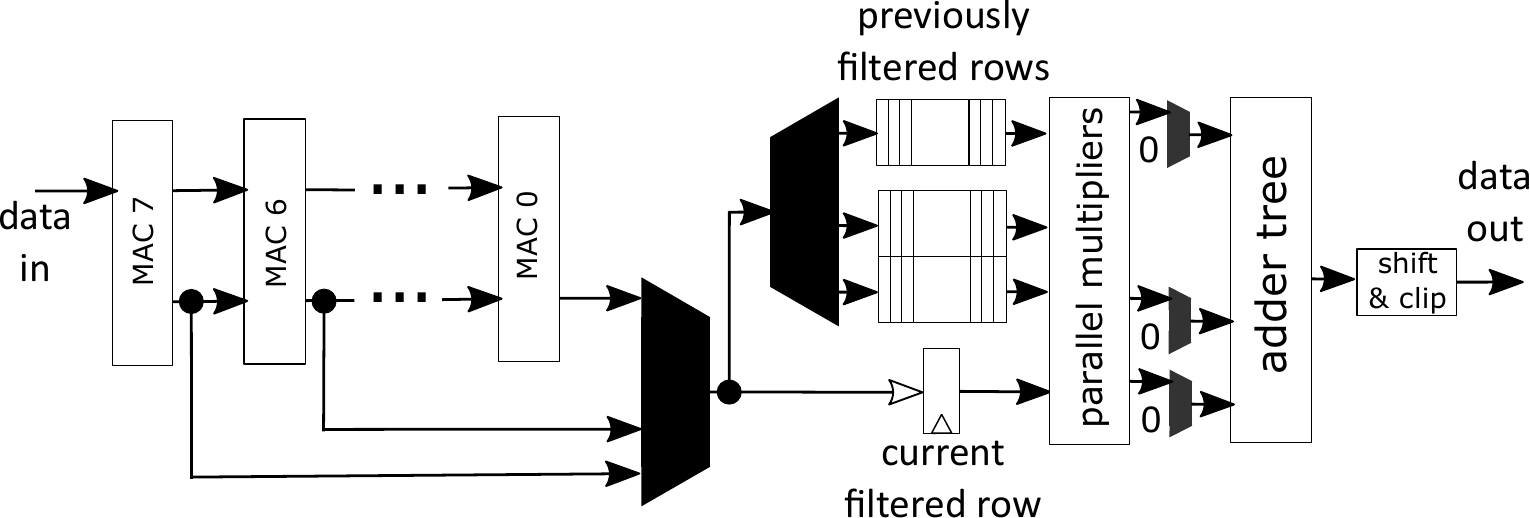}
\caption{Hardware architecture of the reconfigurable approximate HEVC interpolation filter.}
\label{fig:reconf}
\end{figure}

According to the work proposed by Nogues et al. \cite{nogues2015modified}, in \cite{palumbo2015runtime} and \cite{SauPPHNMMR17} the corresponding hardware implementation has been presented. The implementation exploits the full parallelism of the algorithm by pipelining horizontal and vertical filtering by through intermediate FIFO buffers. Virtual coarse-grain reconfiguration (like the one used in \cite{PalumboCPMR14}) is implemented with multiplexers, and allows the system to exclude certain filtering stages from the computation upon request at run-time, thus achieving approximation, as shown in Figure~\ref{fig:reconf}. According to the supported configurations, three different operating points are provided: \emph{H}, where the legacy filters (8/7-tap \emph{luma} and 4-tap \emph{chroma}) are executed; medium, \emph{M}, with 5-tap \emph{luma} and 3-tap \emph{chroma}; low, \emph{L}, with 3-tap \emph{luma} and 2-tap \emph{chroma}. Considering the legacy \emph{H} mode, approximation allows the saving of up to 19\% and 38\% energy per group of frame blocks respectively for luma and chroma. 

At the computational level, approximate computing, due to the reduction of algorithm complexity, can easily lead to an enhancement in terms of throughput. However, the throughput achieved by the legacy (non approximated) implementation could yet be enough for the desired behavior. In such cases, one possibility for a further optimization of the system consumption is to apply frequency scaling on the top of approximation and reconfiguration. Thus, the system throughput can be kept above the minimum (corresponding to the throughput of the legacy configuration), while the power consumption is reduced as a consequence of the frequency decrease. Note that, power consumption reduction through frequency decrease means also execution time enlarging. For this reason, a corresponding energy reduction is not always guaranteed: in \cite{SauPPHNMMR17} the energy saving of under-clocked \emph{L} mode with respect to \emph{H} mode increases from 19\% to 27\% for luma but remains unaltered on 38\% for chroma.

\begin{figure}[t!h]
\centering
\includegraphics[width=\linewidth]{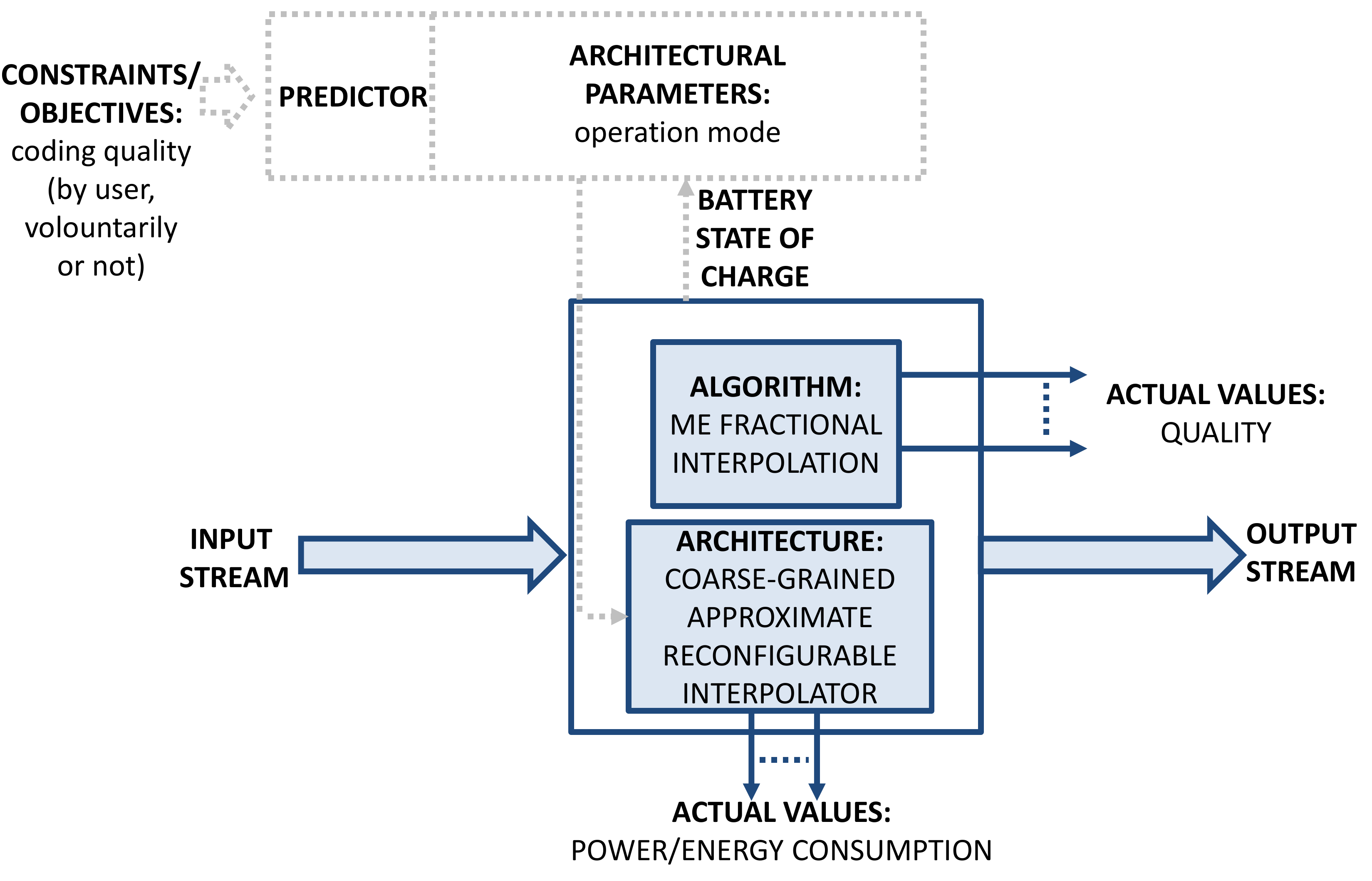}
\caption{Possible dynamic behavior of a reconfigurable approximate interpolator for motion estimation/compensation in HEVC.}
\label{fig:dyn_interp}
\end{figure}

The fractional interpolator for motion estimation proposed in \cite{SauPPHNMMR17}, described as a standalone accelerator without any system level integration, could be used at run-time to play with quality and power/energy parameters. Indeed, as depicted in Figure \ref{fig:dyn_interp}, it could be put in the loop with user or input driven requirement in order to switch among possible architectural operating modes dynamically. Power/energy can be driven by the remaining battery level on the device or by performance monitoring counters providing some kind of information on the current consumption of the system. Quality can be driven by the user, by purpose or automatically depending to the distance of the screen from the human eye. But the operating mode selection, according to the work of Nogues et al. \cite{nogues2015modified}, should also consider the current frame nature (I, P or B) in order to limit the quality degradation, thus introducing some kind of adaptivity in the loop.

To understand the potentials of approximate multi-frequency approaches in the practice, Table \ref{tab:soa} reports a comparison of the work proposed in \cite{SauPPHNMMR17} with similar state of the art solutions. The detailed discussion of such results is beyond the scope of this chapter and it is available in \cite{SauPPHNMMR17}. However, from the table it is possible to appreciate how: i) the power/energy versus quality trade-off curve exists despite the additional logic required to implement reconfiguration on hardware; ii) the reconfigurable approximate multi-frequency solution is capable of competing and beating most of the literature, even not dynamic (not reconfigurable) approaches in terms of throughput and energy efficiency.

\begin{table*}[h!t]
\caption{Potentials of reconfigurable approximate multi-frequency architectures with respect to other literature works. \emph{H}, \emph{M} and \emph{L} are the system operating modes, while in brackets are reported results achievable when multi-frequency is enabled.}
\label{tab:soa}
\centering
\begin{tabular}{cc||c|ccccc|cc|cc}
  & & \rotatebox[origin=c]{90}{\textbf{FPGA, [nm]}} & \rotatebox[origin=c]{90}{\textbf{chroma}} & \rotatebox[origin=c]{90}{\textbf{LUT}} & \rotatebox[origin=c]{90}{\textbf{register}} & \rotatebox[origin=c]{90}{\textbf{BRAM}} & \rotatebox[origin=c]{90}{\textbf{multiplier}} & \rotatebox[origin=c]{90}{\textbf{resolution}} & \rotatebox[origin=c]{90}{\textbf{frame rate [fps]}} & \rotatebox[origin=c]{90}{\textbf{dP [mW]}} & \rotatebox[origin=c]{90}{\textbf{dE x pel [pJ]}} \\
\hline
\hline
\multicolumn{2}{c||}{\emph{\cite{alfonso2013lowcost}}} & Stratix III, 65 & N & 5710 & 3861 & 1 & - & 3840x2160 & 30 & 379 & 1523 \\
\multicolumn{2}{c||}{\emph{\cite{kalali2013reconfigurable}}} & Virtex-6, 40 & N & 1890 & 1224 & 2 & - & 2560x1600 & 60 & 168 & 684 \\
\multicolumn{2}{c||}{\emph{\cite{kalali2014low}}} & Virtex-6, 40 & Y & 3929 & 3422 & 6 & - & 3840x2160 & 30 & NA & NA \\
\multicolumn{2}{c||}{\emph{\cite{diniz2015reconfigurable}}} & Virtex-5, 65 & Y & 5017 & 2550 & 2 & - & 2560x1600 & 30 & 89 & 362 \\
\multicolumn{2}{c||}{\emph{\cite{ghani2016fpga}}} & Virtex-6, 40 & N & 14225 & 9984 & - & - & 3840x2160 & 45 & NA & NA \\
\hline
\multirow{6}{*}{\cite{SauPPHNMMR17}} & \emph{H} & \multirow{3}{*}{Virtex-5, 65} & \multirow{3}{*}{Y} & \multirow{3}{*}{2903} & \multirow{3}{*}{522} & \multirow{3}{*}{28} & \multirow{3}{*}{80} & \multirow{3}{*}{3840x2160} & \multirow{3}{*}{60} & 107(107) & 108(108) \\
 & \emph{M} &  &  &  &  &  &  &  &  & 101(90) & 101(90) \\
& \emph{L} &  &  &  &  &  &  &  &  & 78(64) & 78(64) \\
& \emph{H} & \multirow{3}{*}{Zynq-7000, 28} & \multirow{3}{*}{Y} & \multirow{3}{*}{3235} & \multirow{3}{*}{515} & \multirow{3}{*}{17} & \multirow{3}{*}{80} & \multirow{3}{*}{3840x2160} & \multirow{3}{*}{60} & 64(64) & 64(64) \\
& \emph{M} &  &  &  &  &  &  &  &  & 55(50) & 55(50) \\
& \emph{L} &  &  &  &  &  &  &  &  & 38(32) & 38(32) \\
\end{tabular}
\end{table*}

The work in \cite{MoKS18} combines computational level approximate computing with \emph{Dynamic Voltage and Frequency scaling} (DVFS). Within the approximate computing domain an \emph{imprecise computation task} can be defined as that task that can be logically divided into a \emph{mandatory sub-task} plus an \emph{optional sub-task}. The former has to be completed before the deadline to ensure the baseline Quality of Service (QoS), while the latter may not be completed before the deadline. When the latter situation occurs, QoS associated to the overall task will not be optimal, and different level of approximation can be offered staring from the minimal baseline one. DVFS couples \emph{Dynamic Frequency Scaling} (DFS) and \emph{Dynamic Voltage Scaling} (DVS). DFS relates to the possibility of adjusting the execution frequency of the system on the fly to preserve, according to the computational requirements, the consumed power (which dynamic part depends directly from the frequency) and, in turn, to limit heat dissipation. DVS is the possibility of incrementing (over-volting) or decrementing (under-volting) the reference voltage of a component substantially affecting its power consumption, due to the quadratic relationship between the dynamic power consumption of a given component and its reference voltage. Nevertheless, under-volting negatively affects frequency, which results lowered. This last consideration should make clear why DVS and DFS are strongly related. Indeed, when real-time or user constraints impose quick system response voltage and clock have to be increased together. On the other hand, to minimize power and heat dissipation, they should be minimized whenever possible. Mo et al. in \cite{MoKS18} focus specifically on optimizing task mapping on DVFS-enabled multi-core systems to minimize energy consumption and execution time, while maximizing QoS. This is basically a multi-objective optimization problem, where it is necessary to define: i) in which processor should a certain task (or sub-task) be executed; ii) at which frequency/voltage values (for each processor different couples of frequency-voltage values can be available depending on the underlying multi-core infrastructure). The goal is to maximize the QoS (by completing mandatory part of the computation plus as much optional as possible) while guaranteeing the system constraints (as the available energy budget or the requested real-time). The details of this optimization problem are beyond the scope of this chapter, but \cite{MoKS18} demonstrates how DVFS can be quite effectively used in combination to those applications where imprecise computation task is affordable, and video coding is certainly among them. Basically DVFS is exploited to guarantee enough frequency-voltage levels to those cores running the mandatory parts of the computation, while execution is relaxed for the optional parts according to the achievable compromises among power budget and QoS. In fact, thanks to the various degrees of tolerance behind audio/video streaming, frames with a lower quality are acceptable rather than having any missing frame.  

\subsection{Approximate Computing Key Remarks}\label{ss:APRrem}
In this section the main achievements, potentials and limits of adopting approximate computing within video coding applications are resumed. It has to be remarked that the approximate nature of the same video coding algorithm plays a big role on the achievable results with approximate computing, since it means basically changing the degree of approximation that is already present on the application.

In general, the main considerations that can be derived by observing the state of the art are:
\begin{itemize}
    \item Approximate computing is very effective on the video coding domain, allowing the optimization of throughput, consumption and memory demand. 
    \item Approximation has to be applied selectively and requires a dynamic behavior according to current user requirements/constraints or to current input data. Table \ref{tab:APRclass} resumes the analyzed approximated dynamic solutions.
    \item Frequency-Voltage couple, multi-frequency and multi-VDD approaches seem to be good allies, both in custom heterogeneous accelerators and in commercial multicore infrastructures, to act on approximation involving precision versus power budget. 
    \item In most of the cases, there is still lot of place to further approximate the algorithms and the architectures under different levels of approximations and on several algorithms steps and architecture aspects.
\end{itemize}

\begin{table}[!t]
\caption{Summary of the dynamic (run-time) approximate video coding solutions.}
\label{tab:APRclass}
\centering
\begin{tabular}{r||c|c|c|c|c|c}
\multirow{2}{*}{\textbf{work}} & \multirow{2}{*}{\emph{Algorithm}} & \multirow{2}{*}{\emph{Architecture}} & \emph{Functional} & \emph{Architectural} & \emph{Approx}* & \multirow{2}{*}{\emph{Features}**} \\ 
& & & \emph{Param} & \emph{Param} & \emph{Level} & \\
\hline 
\hline
\multirow{3}{*}{\cite{PalominoSSH16}} & \multirow{3}{*}{HEVC} & \multirow{3}{*}{CPU} & \multirow{3}{*}{operating mode} & - & \multirow{3}{*}{d} & dat, \\
& & & &  & & ndt, \\
& & & &  & & fgr \\ \hline
\multirow{3}{*}{\cite{elHarouni2017}} & \multirow{3}{*}{ME(HEVC)} & \multirow{3}{*}{ASIC/FPGA} & - & tiles size, & \multirow{3}{*}{d, h} & cmp, \\
& & & & tiles approx, & & ndt, \\
& & & & approx bits & & cgr \\ \hline
\multirow{3}{*}{\cite{masera2017}} & \multirow{3}{*}{DCT(HEVC)} & \multirow{3}{*}{ASIC} & - & operating mode, & c & cmp \\
& & & & skip current & & ndt, \\
& & & & rotation & & fgr \\ \hline
\multirow{3}{*}{\cite{nogues2015modified}} & \multirow{3}{*}{HEVC} & \multirow{3}{*}{CPU} & \multirow{3}{*}{operating mode} & - & \multirow{3}{*}{c} & cmp, \\
& & & &  & & ndt, \\
& & & &  & & cgr \\ \hline
\multirow{3}{*}{\cite{SauPPHNMMR17}} & \multirow{3}{*}{ME(HEVC)} & \multirow{3}{*}{FPGA} & - & \multirow{3}{*}{operating mode} & \multirow{3}{*}{c} & cmp, \\
& & & &  & & ndt, \\
& & & &  & & cgr \\ \hline
\multicolumn{7}{c}{* d=data, h=hardware, c=computational-} \\
\multicolumn{7}{c}{** cmp=computation, dat=data; det=deterministic, ndt=non-deterministic;} \\
\multicolumn{7}{c}{cgr=coarse-grained, fgr=fine-grained.}
\end{tabular}
\end{table}

\section{Final Remarks}\label{sec:fin}
The main focus of this chapter was flexibility, intended as the capability of the system to offer different trade-offs reacting to internal and external constraints and triggers. Adaptivity and reconfigurability concepts in Video Coding are not new, and MPEG has studied this concept in various ways. To favor long term adaptivity, that is the capability of better supporting standards evolution, the MPEG Reconfigurable Video Coding (RVC) framework has been defined. MPEG RVC is a dataflow-based codecs specification, which main advantages are: \emph{i)} the possibility of exploiting modularity for parallelization and re-usability purposes, and \emph{ii)} the availability of a wide set of design-time tools helping designers in porting specifications over various targets. RVC-based approaches are used sometimes also at run-time to manage variable constraints, by allowing simultaneous support of multiple codecs, profiles and trade-offs. 

At the state of the art a recent trend to support various trade-off executions on the same system, which is taking place also in the video coding application field, is related to the possibility of adopting approximate computing techniques. Different chapters of this book address this topic, here the emphasis is mainly on how those techniques can be used at run-time, by acting both on functional and architectural system parameters that, by modifying on the fly processing and platform features, enable the possibility of serving variable workload, guaranteeing different requirements and constraints, and of offering different QoS levels. Despite the big promises demonstrated by the presented analysis, approximate computing is not mainstream. In this regard, Ceze and Sampson in \cite{CezeS16} have identified different directions and open issues: investing on \emph{common definitions} and \emph{way of measuring} the quality, \emph{defining proper accelerators} that can aggregate and combine different approximation methodologies, and the \emph{lack of tools support}. The lesson learned in the context of MPEG-RVC may turn out to be useful to address them: raising the level of abstraction along with providing proper design instruments facilitated both the possibility of defining reconfigurable codec specifications and their physical implementation. The usage of co-design techniques, having proper hardware-software interfaces (masking within easily programmable modular hardware accelerators the technological details providing the trade-offs) and a solid compiler workflow, would really unlock the potentials of approximate computing.

\bibliographystyle{spmpsci}

\end{document}